\documentclass[a4paper,11pt]{article}
\usepackage{jcappub} 
\usepackage{amsmath,amssymb,mathtools,bm}
\usepackage{graphicx, color, hepunits}
\usepackage[dvipsnames]{xcolor}
\usepackage{float}
\usepackage{filecontents}
\usepackage{multirow}
\usepackage[english]{babel}

\usepackage{hyperref}
\hypersetup{
     colorlinks   = true,
     citecolor    = red,
	 linkcolor=blue
}
\usepackage{soul}
\let\vec\mathbf

\usepackage{overpic, epsfig, graphicx, url, color, float}
\usepackage[export]{adjustbox}

\usepackage[lofdepth,lotdepth,caption=false,justification=RaggedRight]{subfig}

\newcommand{\es}[2] {\begin{equation} \label{#1} \begin{split} #2 \end{split} \end{equation}}

\newcommand{\mpl}{M_{\rm pl}}
\newcommand{\D}{{\rm d}}
\newcommand{\rhorin}{\rho_{r, \rm in}}
\newcommand{\rhomin}{\rho_{m, \rm in}}

\begin{document}

\title{\LARGE{Towards a Complete Treatment of Scalar-induced Gravitational Waves with Early Matter Domination}}

\author[a]{Soubhik Kumar}
\affiliation[a]{Center for Cosmology and Particle Physics, Department of Physics,
New York University, New York, NY 10003, USA}
\emailAdd{soubhik.kumar@nyu.edu}

\author[b,c]{Hanwen Tai}
\affiliation[b]{Department of Physics, University of Chicago, Chicago, IL 60637, USA}
\emailAdd{hanwentai@uchicago.edu}
\author[b,c,d]{Lian-Tao Wang}

\affiliation[c]{Enrico Fermi Institute, University of Chicago, Chicago, IL 60637, USA}
\affiliation[d]{Kavli Institute for Cosmological Physics, University of Chicago, Chicago, IL 60637, USA}
\emailAdd{liantaow@uchicago.edu}

\abstract{
Large curvature perturbations can source an observable amount of stochastic gravitational wave background (SGWB). 
We consider several scenarios where small-scale curvature perturbations are naturally enhanced due to the presence of additional spectator fields during inflation. 
The same spectator fields can lead to a period of early matter domination (EMD) after inflation. 
We compute the inflationary spectrum of curvature perturbation and determine its evolution at later times, taking into account both the onset and the end of the EMD epoch, and also the impact of relative velocity perturbation between matter and radiation.
The feature that the same field is responsible for both enhanced perturbations and the EMD era, leads to a predictive framework within which the full frequency dependence of SGWB can be computed. 
The SGWB can be observed in several detectors, including those focused on the nano-Hz regime.
Our numerical framework can also be used to study other non-standard cosmological histories.}

\maketitle

\section{Introduction}
Precise measurements of the cosmic microwave background (CMB) and large-scale structure (LSS) have led to a determination of various properties of primordial curvature perturbation at length scales $\gtrsim {\rm Mpc}$~\cite{Planck:2018jri, Bird:2010mp, Chabanier:2019eai}.
At slightly smaller scales, $\sim 0.01-0.1$~Mpc, several recent results have put strong constraints, coming from dwarf galaxies~\cite{Esteban:2023xpk, Dekker:2024nkb} and gravitational lensing~\cite{Gilman:2021gkj}.
On scales even smaller than that, properties of curvature perturbations remain more uncertain.
Some of the constraints in this regime come from CMB $\mu$- and $y$-spectral distortions~\cite{Fixsen:1996nj, 1994ApJ...420..439M, Chluba:2012gq, Chluba:2012we}, stellar streams~\cite{Banik:2019cza, Ando:2022tpj}, Big Bang Nucleosynthesis~\cite{Inomata:2016uip}, heating in ultra-faint dwarf galaxies~\cite{Graham:2024hah}, non-observation of scalar-induced gravitational waves in pulsar timing arrays (PTA)~\cite{EPTA:2015qep}, with the recent PTA results from
NANOGrav~\cite{NANOGrav:2023gor}, EPTA~\cite{EPTA:2023sfo}, EPTA \& InPTA~\cite{EPTA:2023fyk}, PPTA~\cite{Reardon:2023gzh, Zic:2023gta}, CPTA~\cite{Xu:2023wog}  particularly relevant.
Going forward, LSS~\cite{Chung:2023syw}, 21-cm surveys~\cite{Sekiguchi:2013lma, deKruijf:2024voc}, astrometry techniques~\cite{VanTilburg:2018ykj}, direct PTA measurements~\cite{Lee:2020wfn}, and fast radio bursts~\cite{Xiao:2024qay} could significantly improve our knowledge of curvature perturbation up to $\sim 10^{-8}$~Mpc scales.
Excitingly, gravitational wave detectors are sensitive to even smaller scales.

General Relativity predicts that scalar curvature perturbations source tensor perturbations at second order in perturbation theory~\cite{Ananda:2006af, Baumann:2007zm}; see~\cite{Domenech:2021ztg} for an extensive review. 
These tensor perturbations manifest as a stochastic gravitational wave background (SGWB).
Roughly, the abundance of SGWB generated during radiation domination (RD) can be determined as $\Omega_{\rm GW} \sim \Omega_{\rm r} \times (\delta\rho/\rho)^4$, where $\Omega_{\rm r}$ is the present-day energy density in radiation.
Therefore, if the curvature perturbation $\delta\rho/\rho$ is enhanced beyond the $\Lambda$CDM value $\sim 5 \times 10^{-5}$, the sourced SGWB can be observable.\footnote{We give a gauge invariant expression for the curvature perturbation in the next section.}
For example, the upcoming Laser Interferometer Space Antenna (LISA) or the Square Kilometer Array (SKA) might be able to probe $\delta\rho/\rho \sim 10^{-3}-10^{-2}$, see e.g.~\cite{Inomata:2018epa}.
Conversely, a non-observation of such SGWB would lead to novel constraints on $\delta\rho/\rho$ at these extremely small length scales.

 A natural question in this regard is how would $\delta\rho/\rho$ be enhanced than the $\Lambda$CDM value.
 This question has received a significant amount of attention, especially during the last decade, both in the context of SGWB and primordial black holes (PBH).
 In single-field inflationary models, such enhancements could come in several ways; for reviews see~\cite{Green:2020jor, Carr:2020xqk, Ozsoy:2023ryl}.
 Examples of this include a very flat part of the inflaton potential leading to enhanced density perturbations, see e.g.,~\cite{Ivanov:1994pa, Garcia-Bellido:2017mdw, Ballesteros:2017fsr, Hertzberg:2017dkh}; time-varying inflaton mass~\cite{Stewart:1996ey, Stewart:1997wg, Leach:2000ea, Kohri:2007qn}, hill-top models~\cite{Alabidi:2009bk}, among others.
 However, such inflationary potentials may not always be natural (see~\cite{Hertzberg:2017dkh} for an estimate of fine-tuning) and additional ingredients could be necessary to obtain a complete, non-fine-tuned description of inflation and reheating.
 For example, mechanisms to obtain a naturally flat inflaton potential, relevant for CMB-scales, have been vigorously studied to solve the so-called `eta problem', see e.g.,~\cite{Baumann:2014nda}.
 
 Given these considerations, in this work, we focus on a simple class of inflationary models where there is an additional light scalar field along with the inflaton field~\cite{Kasuya:2009up, Kawasaki:2012wr, Kawasaki:2013xsa, Ando:2018nge, Ebadi:2023xhq, Inomata:2023drn}.
 The additional field does not drive inflation and instead acts as a spectator.
 However, it does obtain quantum mechanical fluctuations which can naturally make its perturbation spectrum blue-tilted, implying larger perturbations at smaller length scales.
 During inflation, these are isocurvature perturbations.
 After the end of inflation, the spectator field dilutes less slowly compared to the inflaton decay products which dilute as radiation.
 Consequently, the relative energy density in the spectator field increases with time and eventually, this could lead to a period of early matter domination (EMD).
Around this epoch the primordial (isocurvature) fluctuations of the spectator field no longer act as `spectators', but rather directly determine the curvature perturbation.
The originally blue-tilted isocurvature spectrum thus leads to a blue-tilted curvature perturbation.
At length scales $\gtrsim$ Mpc, this blue-tilted contribution is small enough and the slightly red-tilted inflaton contribution dominates, reproducing the CMB and LSS observations.

A generic consequence of this cosmology is a broad and gradual rise of the primordial curvature power spectrum as we go to smaller length scales.
This is to be compared with a sharper rise of the spectrum, common in single-field scenarios~\cite{Ivanov:1994pa, Garcia-Bellido:2017mdw, Ballesteros:2017fsr, Hertzberg:2017dkh}.
In fact, as we will show later, in certain scenarios the enhanced power spectrum at small length scales can be approximately flat over a range of scales, and therefore the associated SGWB could be observable in multiple GW detectors operating at different frequencies.
A simultaneous observation of SGWB at vastly different frequencies would give a powerful indication of the primordial origin of the signal.
Notably, these scenarios do not require any specific shape or fine-tuning of the inflaton potential.

The spectator field could naturally lead to a period of EMD, as we described above.
However, this complicates the determination of the strength of SGWB.
While there exist analytical derivations of SGWB from curvature perturbation when the equation of state $w$ and the speed of propagation of fluctuations $c_s$ is constant~\cite{Espinosa:2018eve, Kohri:2018awv, Domenech:2019quo}, in our situation of interest these two parameters are time dependent.
This is because we are interested {\it both} in the onset of EMD and the end of EMD.
As a result, the sub-horizon evolution for Newtonian potential needs to be determined numerically for each individual $k$-mode.
To this end, we analytically derive the appropriate initial conditions for the superhorizon modes, tracking the subdominant matter abundance at the initial time.
Then we adopt a fully numerical approach and solve the Boltzmann equation that couples (decaying) matter, radiation, and metric fluctuations.
We then integrate the resulting Newtonian potential with the appropriate Green function, which we also determine numerically, to obtain the final frequency spectrum of GW.
In the process, we also track the contribution from relative velocity perturbation between matter and radiation, which has a significant impact for certain scales or frequencies.
We emphasize that our numerical framework capturing the start and the end of EMD is independent of the precise form of the curvature spectrum and can be used even when there are sharp peaks in the spectrum, as in single-field scenarios.
We provide a {\tt Mathematica} notebook~\cite{SIGW_EMD} which contains integration kernels that encode the full numerical evolution of modes.
These kernels can be used straightforwardly to evaluate the strength of the SGWB, as long as homogeneous cosmology is the same as described in this work, but for arbitrary primordial power spectrum.

As we will explain below, the frequency spectrum of SGWB could exhibit both a broken power-law peak or a flattened peak.
The precise location of these peaks can be calculated from first principles and is dependent on the inflationary Hubble scale and the mass and lifetime of the spectator field, among other parameters.
This is because the same spectator field leads to an enhanced curvature spectrum and the epoch of EMD.
We will show that depending on the parameters, a variety of GW detectors probing nano-Hz to deci-Hz frequencies are sensitive to these scenarios.
In particular, we show example benchmarks where future runs of the PTA measurements of the NANOGrav~\cite{NANOGrav:2023gor}, EPTA~\cite{EPTA:2023sfo}, EPTA \& InPTA~\cite{EPTA:2023fyk}, PPTA~\cite{Reardon:2023gzh, Zic:2023gta}, CPTA~\cite{Xu:2023wog} collaborations in the nano-Hz regime would constitute powerful probes. 
We show one example where we compare our prediction with the NANOGrav observation~\cite{NANOGrav:2023gor, NANOGrav:2023hvm}.

{\bf Comparison with the literature. }
The impact of an EMD era on scalar-induced SGWB has a long history~\cite{Assadullahi:2009nf, Alabidi:2013lya, Kohri:2018awv, Inomata:2019zqy, Inomata:2019ivs, Pearce:2023kxp}.
Ref.~\cite{Assadullahi:2009nf} worked in the limit of a pure matter-dominated era with $c_s^2=w=0$, while Ref.~\cite{Alabidi:2013lya} employed some approximate expressions for the scale factor and SGWB source term to describe the transition between an EMD and radiation-dominated era.
Subsequently, Ref.~\cite{Kohri:2018awv} also focused on the transition to and from an EMD era, obtaining useful semianalytic results.
This was revisited by Refs.~\cite{Inomata:2019zqy, Inomata:2019ivs} which took into account the Boltzmann equations coupling decaying matter, radiation, and the Newtonian potential $\Phi$.
They also approximated the time-evolution of $\Phi$ by a {\it $k$-independent} fitting formula, instead of assuming it remains constant till the time of EMD-radiation equality, as done in the previous works.
However, usage of this fitting formula required imposing an artificial lower cutoff on $k$, below which the fitting formula was not a good approximation.
Nonetheless, with these improvements, Ref.~\cite{Inomata:2019zqy} showed that for gradual transitions, as relevant for particle decays, the SGWB strength is much smaller than what was found in Ref.~\cite{Kohri:2018awv}.
In reality, however, the time evolution of $\Phi$ is different for each $k$-mode, although taking that into account requires intensive computation.
Ref.~\cite{Pearce:2023kxp} performed one such computation by tracking each individual $k$-mode (with some lattice spacing) to evaluate the final SGWB spectrum.
They focused on the end of the EMD epoch, as the Universe transitions from EMD to RD.

In this work, we improve on this previous literature in several ways. First, we track the full evolution of each $k$-mode as was done in~\cite{Pearce:2023kxp}, but we consider {\it both} the onset and the end of the EMD epoch in realistic cosmologies.
This requires deriving a new set of initial conditions for superhorizon perturbations.
Additionally, this has important observational consequences since the SGWB generated prior to the onset of EMD would get diluted when EMD ends, due to entropy injection from matter decay. This leads to a suppression of the high-frequency part of the SGWB spectrum.
Second, we focus on scenarios where we can compute the full spectrum of curvature perturbation given the inflationary dynamics and therefore we do not need to impose any hard cutoff for large $k$ as in~\cite{Inomata:2019zqy, Pearce:2023kxp} and we can also include the effect of a significant blue tilt.
Because of the same reason, we can obtain quantitative predictions for the strength SGWB without treating $\delta\rho/\rho$ as a free parameter.
Third, we show that the relative velocity perturbation between matter and radiation contributes to SGWB in a significant way.
While this term was discussed in the context of primordial black hole domination~\cite{Domenech:2020ssp} and investigation of gauge invariance of SGWB~\cite{Gurian:2021rfv}, it was concluded that the effect is subdominant in those cases.
In our scenario, where EMD ends due to particle decay, the effect is important and we take this into account for the first time. 

The rest of this work is organized as follows.
In Sec.~\ref{sec:gen} we describe the general form of the curvature power spectrum in the inflationary scenarios of interest.
Then in Sec.~\ref{sec:mech} we describe three related classes of mechanisms that can lead to blue-tilted (iso)curvature perturbations.
We show benchmark examples and compute the shape of the full curvature power spectrum in Sec.~\ref{sec:spectrum}.
We then outline in Sec.~\ref{sec:comp} the formalism for computing the strength of SGWB by taking into account the EMD epoch, providing the explicit computation in Appendix~\ref{sec:details}.
We also illustrate the impact of the relative velocity term discussed above.
Using these in Sec.~\ref{sec:detectors} we show the predictions of SGWB for a variety of GW detectors.
We conclude in Sec.~\ref{sec:conc}.

\section{Generalities}
\label{sec:gen}
We consider scenarios with two light, dynamical scalar fields that obtain large-scale primordial fluctuations during inflation: the inflaton $\phi$, and a spectator field $\chi$.
After inflation, $\phi$ reheats into the Standard Model (SM) radiation bath, decoupled from $\chi$.
On the other hand, $\chi$ remains frozen in its potential until the Hubble scale falls below its (effective) mass.
Following this, $\chi$ oscillates in its potential and eventually starts diluting like matter.
We assume $\chi$ is sufficiently long-lived and gives rise to an epoch of EMD when its energy density dominates the total energy density of the Universe.
Finally, $\chi$ decays into SM radiation, following which the evolution of the Universe is as in the standard $\Lambda$CDM cosmology.

To describe the density fluctuations with the above cosmological history, we use the curvature perturbation on uniform density hypersurfaces defined as~\cite{Malik:2008im},
\es{}{
\zeta = -\Psi - H {\delta\rho \over \dot{\bar{\rho}}},
}
where the scalar perturbations of the metric are defined in the Newtonian gauge as,
\es{eq:newton0}{
\D s^2 = a^2(\eta)\left( -(1+2\Phi)\D \eta^2 + (1-2\Psi)\D \vec{x}^2\right).
}
The Hubble rate is denoted by $H\equiv \dot{a}(t)/a(t)$, the density perturbation by $\delta\rho$, and the homogeneous energy density as $\bar{\rho}$. Here and below an overdot (prime) denotes derivatives with respect to cosmic time $t$ (conformal time $\eta$).
Unless mentioned otherwise, throughout this work the symbol $H$ will denote the Hubble scale during inflation, which we take to be approximately constant.

We can write $\zeta$ in terms of gauge invariant fluctuations of the individual components~\cite{Malik:2008im}:
\es{}{
\zeta = {\dot{\bar{\rho}}_r \over \dot{\bar{\rho}}_r + \dot{\bar{\rho}}_\chi} \zeta_r + {\dot{\bar{\rho}}_\chi \over \dot{\bar{\rho}}_r + \dot{\bar{\rho}}_\chi} \zeta_\chi = {4 \bar{\rho}_r \over 4 \bar{\rho}_r + 3\bar{\rho}_\chi}\zeta_r + {3 \bar{\rho}_\chi \over 4 \bar{\rho}_r + 3\bar{\rho}_\chi}\zeta_\chi.
}
Here $\zeta_i = -\Psi -H \delta\rho_i/\dot{\bar{\rho}}_i$ for $i=r$ (radiation bath from $\phi$ decay) and $\chi$.
In the second equality, we have used the standard redshift of radiation and matter, $\dot{\bar{\rho}}_r=-4H\bar{\rho}_r$ and $\dot{\bar{\rho}}_\chi=-3H\bar{\rho}_\chi$, valid before $\chi$ decay.
We can rewrite this in terms of isocurvature perturbation $S_\chi\equiv3(\zeta_\chi-\zeta_r)$,
\es{eq:zeta_tot}{
\zeta = \zeta_r + {\bar{\rho}_\chi \over 4 \bar{\rho}_r + 3 \bar{\rho}_\chi} S_\chi.
}
The fluctuations of $\chi$ are uncorrelated with $\phi$, and hence the radiation bath $r$.
Therefore, the total power spectrum is a sum of the two individual contributions, without any correlation,
\es{eq:power_tot}{
\Delta_\zeta^2 = \Delta_{\zeta_r}^2 + \left({\bar{\rho}_\chi \over 4 \bar{\rho}_r + 3 \bar{\rho}_\chi}\right)^2 \Delta_{S_\chi}^2,
}
where $\Delta_\zeta^2(k) \equiv k^3\langle\zeta(\vec{k})\zeta(-\vec{k})\rangle/(2\pi^2)$ is defined in terms of the two-point function of curvature perturbation.
The quantities $\Delta_{\zeta_r}^2$ and $\Delta_{S_\chi}^2$ are defined analogously.

Prior to the decay of $\chi$, the coefficient multiplying $S_\chi$ in Eq.~\eqref{eq:zeta_tot} is time-dependent and perturbations are not adiabatic.
This means $\Delta_\zeta^2(k)$ changes with time, even on superhorizon scales.
This can be described by writing,
\es{eq:power_tot}{
{\bar{\rho}_\chi(t) \over 4 \bar{\rho}_r(t) + 3 \bar{\rho}_\chi(t)} = {\bar{\rho}_\chi(t_{\rm d})(a(t)/a(t_{\rm d})) \over 4 \bar{\rho}_r(t_{\rm d}) + 3 \bar{\rho}_\chi(t_{\rm d})(a(t)/a(t_{\rm d}))},
}
where $t_{\rm d}$ is the time of $\chi$-decay.
After $\chi$ decays into radiation, $\zeta$ remains constant with time on superhorizon scales.
We consider scenarios where $\chi$ dominates the energy density of the Universe before its decay, giving rise to a period of EMD.
Approximating both the decay process and any prior transition from radiation domination (RD) to EMD to be instantaneous,\footnote{In our numerical computation below we do not make this approximation.} we obtain
\es{eq:Delta_zeta}{
\Delta_\zeta^2(k) =
\begin{cases}
     \Delta_{\zeta_r}^2(k) + \left(r_{\rm d} \over 4+3 r_{\rm d}\right)^2 \Delta_{S_\chi}^2(k) & \text{for } k_{\rm d} \geq k, \\
   \Delta_{\zeta_r}^2(k) + \left({r_{\rm d} (k_{\rm d}/k)^2 \over 4  + 3 r_{\rm d}(k_{\rm d}/k)^2}\right)^2  \Delta_{S_\chi}^2(k) & \text{for } k_{\rm EMD} \geq k > k_{\rm d}, \\
   \Delta_{\zeta_r}^2(k) + \left({r_{\rm d} (k_{\rm d}^2/(k_{\rm EMD} k)) \over 4 + 3 r_{\rm d}(k_{\rm d}^2/(k_{\rm EMD}k))}\right)^2  \Delta_{S_\chi}^2(k) & \text{for } k > k_{\rm EMD}.
\end{cases}
}
Here
\es{eq:Rd}{
r_{\rm d} \equiv {\bar{\rho}_\chi(t_{\rm d}) \over \bar{\rho}_r(t_{\rm d})},
}
is a constant that determines the infrared (IR) part of the spectrum.
The modes $k_{\rm d}$ and $k_{\rm EMD}$ reenter the horizon at the time of $\chi$-decay and onset of matter domination induced by $\chi$, respectively: $k_{\rm d} = a(t_{\rm d}) H(t_{\rm d})$ and $k_{\rm EMD} = a(t_{\rm EMD}) H(t_{\rm EMD})$.
Therefore, both the onset of EMD and $\chi$-decay generate non-trivial scale dependences in the power spectrum.
This is in addition to the intrinsic scale dependence of $\Delta_{S_\chi}^2$ that will be discussed in Secs.~\ref{sec:mech} and~\ref{sec:spectrum}.

We require that $\Delta_\zeta^2$ obeys the constraints from CMB and structure of matter distribution on relatively larger scales $\gtrsim 0.1~{\rm Mpc}$.
As we will see in Sec.~\ref{sec:mech}, the power spectrum $\Delta_{S_\chi}^2$ is generically blue-tilted.
Therefore, to be consistent with the observed slightly red-tilted spectrum at those large scales, the contribution to $\Delta_\zeta^2$ from $\Delta_{S_\chi}^2$ should be sufficiently small, and $\Delta_\zeta^2$ should instead be primarily determined by the inflaton contribution, $\Delta_{\zeta_r}^2$.
At much smaller scales, however, due to the blue tilt, $\Delta_{S_\chi}^2 \gg \Delta_{\zeta_r}^2$.
Therefore, the leading contribution to $\Delta_\zeta^2$ comes from $\Delta_{S_\chi}^2$, especially for $r_{\rm d} \sim 1$.
We can summarize all these constraints by computing the tilt $n_s$ and the running of the tilt $\D n_s/\D \ln k$:
\es{}{
n_s-1 = {\D \ln \Delta_\zeta^2 \over \D \ln k} &= {\Delta_{\zeta_r}^2 \over \Delta_\zeta^2} {\D \ln \Delta_{\zeta_r}^2 \over \D \ln k} + \left(r_{\rm d} \over 4+3 r_{\rm d}\right)^2 {\Delta_{S_\chi}^2 \over \Delta_\zeta^2} {\D \ln \Delta_{S_\chi}^2 \over \D \ln k},\\
{\D n_s \over \D \ln k} &\approx \left(r_{\rm d} \over 4+3 r_{\rm d}\right)^2 {\Delta_{S_\chi}^2 \over \Delta_\zeta^2} \left({\D \ln \Delta_{S_\chi}^2 \over \D \ln k}\right)^2.
}
Here we have considered $k < k_{\rm d}$, as relevant for CMB and structure constraints, and given our benchmark choices for $k_{\rm d}$ (to be described in Sec.~\ref{sec:spectrum}).
We also consider scenarios where the tilt of  $\Delta_{S_\chi}^2$ is much larger (in magnitude) than that of $\Delta_{\zeta_r}^2$, and therefore, we have ignored the running of the tilt of both $\Delta_{\zeta_r}^2$ and $\Delta_{S_\chi}^2$.

\section{Mechanisms}\label{sec:mech}
Having discussed the general aspects of density perturbations relevant to this work, we now focus on mechanisms that can give rise to a blue-tilted spectrum of $\Delta_{S_\chi}^2$.
\subsection{Scenario 1: Stochastic Curvaton}\label{sec:s1}
The first scenario corresponds to quantum fluctuations of a light scalar field around the bottom of its potential~\cite{Starobinsky:1986fx, Starobinsky:1994bd, Linde:1996gt}.
As shown in~\cite{Ebadi:2023xhq}, these fluctuations can naturally have a blue-tilted spectrum to give rise to an observable SGWB.
The idea can be illustrated with a simple model,
\es{eq:Vchi}{
V(\chi) = {1\over 2}m^2 \chi^2 + {1\over 4}\lambda \chi^4.
}
The dynamics of a light field such as $\chi$, with $|V''(\chi)| \ll H^2$, can be described by a Langevin equation~\cite{Starobinsky:1986fx, Starobinsky:1994bd, Sasaki:1987gy, Nambu:1987ef},
\es{}{
\dot{\chi}(t,\vec{x}) = -{1\over 3H}V'(\chi) + \xi(t,\vec{x}).
}
The first term on the right-hand side represents a classical motion of $\chi(t,\vec{x})$ on its potential $V(\chi)$, while the second term $\xi$ captures `noise' from quantum fluctuations of subhorizon modes.
The latter satisfies a correlation function,
\es{}{
\langle \xi(t,\vec{x}) \xi(t',\vec{x}')\rangle = {H^3 \over 4\pi^2}\delta(t-t')j_0(\varepsilon a H|\vec{x}-\vec{x}'|),
}
with $j_0$ the zeroth-order Bessel function and $\varepsilon$ a numerical factor $\sim 1$.
Physically, these quantum noise fluctuations contribute to a random-walk behavior of the classical field $\chi(t,\vec{x})$.
Over a long time scale, these fluctuations get accumulated, and as inflation goes on, new noise fluctuations get produced and eventually get stretched to a size larger than the Hubble horizon and become part of the classical field.

The power spectrum of density fluctuations in this scenario can be computed using a Fokker-Planck equation for the probability distribution function of the field $\chi(t,\vec{x})$.
This has been reviewed extensively in the prior literature~\cite{Starobinsky:1986fx, Starobinsky:1994bd, Sasaki:1987gy, Nambu:1987ef, Ebadi:2023xhq, Graham:2018jyp, Markkanen:2019kpv}; the power spectrum can be computed as~\cite{Markkanen:2019kpv},
\es{eq:Delta_S}{
\Delta_{S_\chi}^2(k) = {2\over \pi}\sum_n  g_n^2 \Gamma\left(2-2\Lambda_n/H\right) \sin(\pi \Lambda_n/H)\left(k \over k_{\rm end}\right)^{2\Lambda_n/H},
}
in terms of the $\Gamma$ function and the comoving horizon size at the end of inflation $k_{\rm end}=a_{\rm end} H_{\rm end}$.
Here $\Lambda_n$'s are eigenvalues corresponding to an orthonormal basis of eigenfunctions $\psi_n$ that satisfy the equation
\es{eq:eigen}{
{\partial^2\psi_n(\chi) \over \partial\chi^2} - \left(\tilde{V}'^2 - \tilde{V}''\right)\psi_n(\chi) = -{8\pi^2 \over H^3}\Lambda_n \psi_n(\chi),
}
with $\tilde{V}(\chi) = 4\pi^2 V(\chi)/(3H^4)$.
The quantity $g_n$ can be computed as,
\es{}{
g_n = {\int \D \chi \psi_n(\chi)\bar{\rho}(\chi)\psi_0(\chi)\over \int \D \chi \psi_0(\chi)\bar{\rho}(\chi)\psi_0(\chi)},
}
where $\bar{\rho}(\chi)\approx V(\chi)$ since $\chi$ is frozen on its potential.
This eigenvalue equation can be solved numerically for a given scalar potential $V(\chi)$, as we do later.

We can obtain a physical understanding of the origin of the blue tilt as follows~\cite{Ebadi:2023xhq}.
We assume a negligible quartic coupling $\lambda$ for this discussion.
The classical field $\chi(t,\vec{k})$ then satisfies a `slow-roll' equation $3H\dot{\chi}+m^2\chi \approx 0$ on superhorizon scales which can be approximately solved as,
\es{eq:sr}{
|\chi(t,\vec{k})| \sim {H \over k^{3/2}} \exp\left(-{m^2 (t-t_*) \over 3H}\right).
}
Here $t_*$ is the time of horizon exit of the $k$-mode, $k = a(t_*) H$.
Smaller $k$-modes exit the horizon earlier and experience more damping as per Eq.~\eqref{eq:sr} than larger $k$-modes.
Therefore, we expect a larger power spectrum for larger $k$.
To quantify this we can compare the power in two $k$-modes using Eq.~\eqref{eq:sr} and approximating $a(t) = \exp(H t)$ during inflation,
\es{}{
{k_1^3 |\chi(t,\vec{k}_1)|^2 \over k_2^3 |\chi(t,\vec{k}_2)|^2} \sim \left(k_1\over k_2\right)^{2m^2/(3H^2)}.
}
This shows that the power spectrum is indeed larger at larger $k$ and the blue tilt is determined by $m$.
An analysis using Eq.~\eqref{eq:eigen} also gives the same expression for the tilt~\cite{Markkanen:2019kpv}; see also~\cite{Linde:1996gt}.
In the following, we also include $\lambda$ and compute the eigenvalues $\Lambda_n$ by solving Eq.~\eqref{eq:eigen}, and then obtain $\Delta_{S_\chi}^2(k)$ using Eq.~\eqref{eq:Delta_S}.
That, together with Eq.~\eqref{eq:Delta_zeta}, determines the power spectrum of curvature perturbation.
As an illustration, for $m^2 = 0.2 H^2$ and $\lambda=0.1$, we find a blue tilt $\D(\ln \Delta_{S_\chi}^2)/(\D \ln k)\approx 0.36$~\cite{Ebadi:2023xhq}.

\subsection{Scenario 2: Misaligned Curvaton}
\label{sec:s2}
Blue-tilted perturbations can also arise in scenarios where $\chi$ is `misaligned' around some non-zero field value $\chi_0$, as opposed to fluctuating around the minimum, as considered above.
This can be illustrated with a similar Lagrangian as in Eq.~\eqref{eq:Vchi} with $\lambda=0$.
Through the same argument as above, we find
\es{eq:chi0_rat}{
{\chi_0^2(t_1)\over \chi_0^2(t_2)} = \exp\left(-{2m^2 \over 3H}(t_1-t_2)\right) = \left({k_1\over k_2}\right)^{-2m^2/(3H^2)}.
}
Here $t_1$ and $t_2$ are the horizon-exit time of the modes $k_1$ and $k_2$.
As we will see below, in this scenario $\Delta_{S_\chi}^2 \propto 1/\chi_0^2$, implying a blue-tilted spectrum.

The primary difference between this and the stochastic scenario above lies in the fact that the energy density in $\chi$ during inflation is given by $m^2 \chi_0^2$ which could be much larger than $\sim H^4$, the energy density in $\chi$ in the stochastic scenario.
This implies the epoch of EMD induced by $\chi$ is different in the two scenarios, which has important implications for the frequency dependence of SGWB.
We will explore this in later sections.

The isocurvature fluctuations in this context are defined via,
\es{}{
S_\chi = {\delta \rho_\chi \over \bar{\rho}_\chi} = {\chi^2 -\langle\chi^2\rangle \over \langle\chi^2\rangle} = {2\chi_0 \delta\chi + (\delta\chi)^2 - \langle(\delta\chi)^2\rangle \over \chi_0^2 + \langle(\delta\chi)^2\rangle},
} 
where $\bar{\rho}_\chi \propto \chi^2$ is the energy density in the curvaton and $\langle\chi\rangle = \chi_0$ with $\chi = \chi_0+\delta\chi$.
The power spectrum of $S_\chi$ is then given by~\cite{Hertzberg:2008wr},
\es{eq:Delta_S}{
\Delta_{S_\chi}^2 = \langle S_\chi^2 \rangle = {4 \chi_0^2 \langle(\delta\chi)^2\rangle + 2\langle(\delta\chi)^2\rangle^2 \over \left(\chi_0^2 + \langle(\delta\chi)^2\rangle\right)^2}.
}
In this case we are interested in scenarios with $\chi_0^2 \gg \langle(\delta\chi)^2\rangle$, such that we can approximate,
\es{eq:mis_curv}{
\Delta_{S_\chi}^2 \approx {4\over \chi_0^2} \langle(\delta\chi)^2\rangle = \left({H \over \pi \chi_0}\right)^2.
}
From this expression, we see that any time dependence of $\chi_0$ would lead to a scale dependence of $\Delta_{S_\chi}^2$~\cite{Linde:1996gt, Lyth:2001nq}.
In particular, if $\chi_0$ decreases with time, such as in Eq.~\eqref{eq:chi0_rat}, then $\Delta_{S_\chi}^2$ increases with time.
This implies modes that exit earlier during inflation (smaller $k$) have less power compared to modes that exit later (larger $k$), i.e., the spectrum of $\Delta_{S_\chi}^2$ is blue-tilted.

\subsection{Scenario 3: Rolling Radial Mode}
\label{sec:s3}
A similar situation arises in the presence of one or more global $U(1)$ symmetries.
This class of models has been discussed in previous literature as part of `axion curvaton models'~\cite{Kasuya:2009up, Kawasaki:2012wr, Kawasaki:2013xsa, Ando:2017veq, Ando:2018nge, Kawasaki:2021ycf, Inomata:2023drn}.
The basic idea behind these models is to have $\chi$ as a Goldstone mode from one or more $U(1)$ symmetry breaking.
Similar to Eq.~\eqref{eq:mis_curv} one can then write
\es{}{
\Delta_{S_\chi}^2 \approx  \left({H \over \pi \theta_i \langle s\rangle}\right)^2,
}
where $\langle s\rangle$ is the homogeneous component of the radial mode $s$.
The blue-tilt in this class of models arises from the fact that during inflation $s$ need not stay fixed at its minimum, but rather can roll towards it, given appropriate initial conditions.\footnote{A dynamical `decay constant' can also play a role in determining the dark matter abundance, see e.g.,~\cite{Allali:2022yvx}.}
For example, if $s$ rolls from large field values towards small field values during inflation, modes exiting later during inflation have more power.
That is, the spectrum is blue-tilted.
However, once the field reaches the minimum, $\langle s \rangle$ remains constant with time, implying an almost flat spectrum of $\Delta_{S_\chi}^2$.

We can illustrate this behavior by considering a `Mexican hat' potential for a complex scalar field $\Phi$,
\es{}{
V(\Phi) = \lambda_\Phi (|\Phi|^2 - f_a^2/2)^2 - \left({1\over 4}m^2\Phi^2+{\rm h.c.}\right),
}
with $\Phi = (s/\sqrt{2})\exp(i\theta)$ and $m$ parametrizing a soft-breaking of the $U(1)$ symmetry.
Including the kinetic terms we can write,\footnote{The soft breaking also contributes to a mass term for $s$ which is however subdominant for the our parameter choice described later and is not included here.}
\es{eq:L_mexhat}{
{\cal L} \supset {1\over 2}(\partial_\mu s)^2 + {1 \over 2}s^2 (\partial_\mu\theta)^2 -\lambda_\Phi (s^2 -f_a^2)^2/4 + {1\over 2} m^2 s^2 \theta^2.
}
The Goldstone mode $\theta$ is a light field during inflation and since it has a subdominant energy density during inflation, its fluctuation corresponds to isocurvature modes.
We denote an initial misalignment angle as $\theta_i = \langle \theta\rangle$ and the fluctuations as $\delta\theta = \theta -\theta_i$.
The canonically normalized Goldstone mode is given by $\chi \equiv \langle s\rangle \theta$, assuming the homogeneous VEV $\langle s\rangle$ is slowly varying with respect to the Hubble rate.
The power spectrum of fluctuations is then given by $\langle (\delta\theta)^2\rangle = H^2/(2\pi \langle s\rangle)^2$.
Following the same steps as in the previous subsection, we can derive
\es{eq:iso_power_s3}{
\Delta_{S_\chi}^2 \approx {4\over \theta_i^2} \langle(\delta\theta)^2\rangle = \left({H \over \pi \theta_i \langle s\rangle}\right)^2.
}

\begin{figure}[t]
\centering
\includegraphics[width=0.45\textwidth]{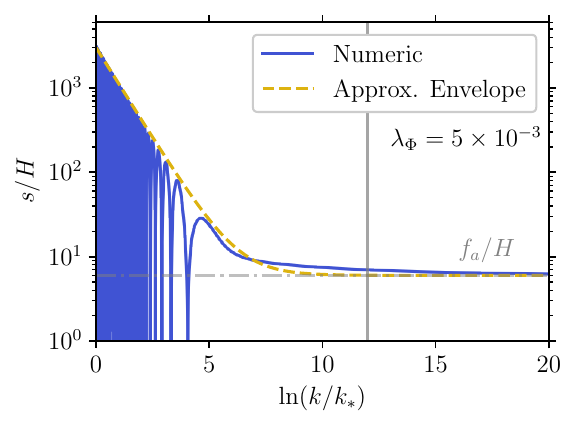}
\includegraphics[width=0.45\textwidth]{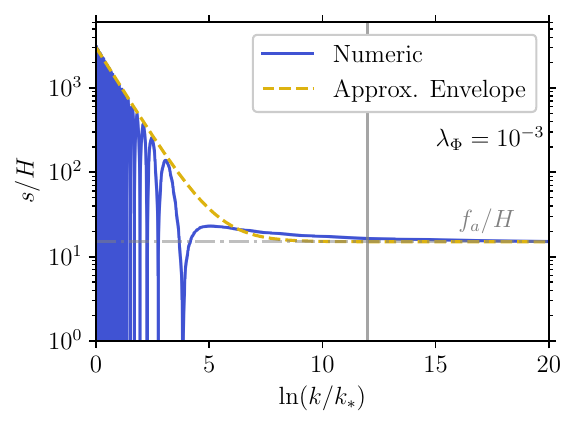}
\caption{The time evolution of the radial mode given Eq.~\eqref{eq:L_mexhat} (solid blue). We parametrize time in terms of the time of horizon exit of a $k$-mode, $k = a H \approx \exp(H t) H$, and a fiducial scale $k_*$. The field starts displaced on its potential and eventually rolls down to $f_a$ (dot-dashed gray). Accordingly, the spectrum of $\Delta_{S_{\chi}}^2$ would become flat approximately for $\ln(k/k_*)>\ln(k_c/k_*)=12$ (shown via the vertical gray lines), for both left and right panels. We show an approximation of the envelope in dashed yellow, by choosing a value of the radial mode mass $m_s$.} 
\label{fig:s_dyna}
\end{figure}

To understand the behavior of $\Delta_{S_\chi}^2$ as a function of scale, we can solve for the homogeneous dynamics of $s$ using Eq.~\eqref{eq:L_mexhat}.
For this purpose, we can neglect the angular mode $\theta$. 
We show the numerical result in Fig.~\ref{fig:s_dyna} which shows that $s$ rolls from its initial location, oscillates around the minimum and eventually settles at $f_a$.
To explicitly show the scales involved, we replace the time coordinate $t$ in terms of the $k$-mode that exits the horizon at that time: $\ln(k/k_*)=H(t-t_*)$.
Here $t_*$, or equivalently $k_*$, is a fiducial time and in particular, we fix $k_* = 20~{\rm Mpc}^{-1}$.
Physically, the scale $k_*$ exits the horizon when $s$ starts to move on its potential.
Its time evolution can approximated by $s \sim \exp(-(3/2-\nu)Ht)$ where $\nu=\sqrt{9/4-m_s^2/H^2}$ and $m_s$ is chosen to fit the envelope of the oscillations in Fig.~\ref{fig:s_dyna}.  

At the scale $k_*$, the correction to $\Delta_\zeta^2$ is small enough compared to the current precision and the energy density in the radial mode is also subdominant compared to the inflationary energy density.
The behavior of $\Delta_\zeta^2$ for $k<k_*$ depends on the dynamics of the radial mode prior to the time $t_*$ and is model dependent.
However, for $k\gg k_*$, $\Delta_{S_\chi}^2 \sim 10^{-2}-10^{-3}$ without violating the current bounds.
For the examples shown in Fig.~\ref{fig:s_dyna}, $s$ settles down to the minimum approximately for $\ln(k/k_*)>12$ implying $\Delta_{S_{\chi}}^2(k)$ is approximately flat for $\ln(k/k_*)>\ln(k_c/k_*)=12$.
Choosing $k_*=20~{\rm Mpc}^{-1}$, this implies a flattening for $k>k_c$ with $k_c \approx 3.2\times 10^6~{\rm Mpc}^{-1}$. 
This will serve as our benchmark example of an almost flat but relatively large $\Delta_{S_{\chi}}^2(k)$.
In Sec.~\ref{sec:spectrum} we will use the notation $\chi_{0,\rm end} \equiv f_a \theta_i$ to denote the normalized field value of the Goldstone mode after $s$ has settled to its minimum.

\section{Spectrum of Curvature Perturbation}\label{sec:spectrum}
\begin{figure}[t]
\centering
\includegraphics[width=0.7\textwidth]{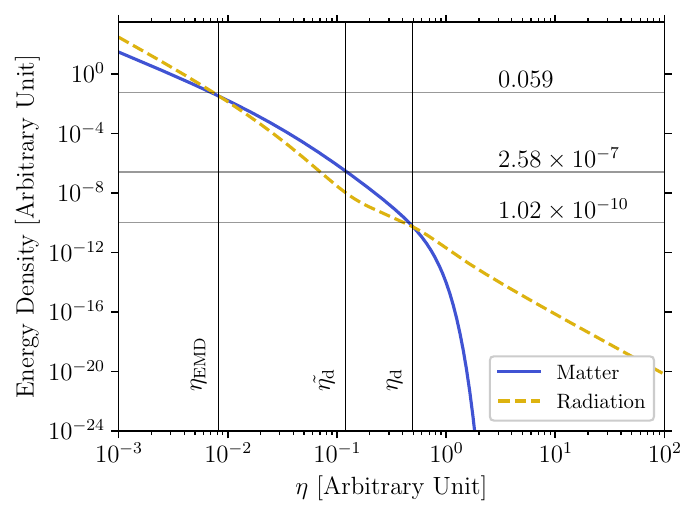}
\caption{Evolution of energy density in (decaying) matter and radiation. The conformal times $\eta_{\rm EMD}$, $\tilde{\eta_{\rm d}}$, and $\eta_{\rm d}$ denote the onset of EMD, the decay time of $\chi$ and the end of EMD, respectively. The total energy densities (sum of matter and radiation) at these three times are $\rho_{\rm EMD}$, $\tilde{\rho}_{\rm d}$, and $\rho_{\rm d}$, respectively, and are indicated by the horizontal lines with the corresponding numerical values. While these absolute values are not relevant, their ratios determine the relevant cosmology, as discussed in the text.}
\label{fig:bg_evol}
\end{figure}

To compute the curvature perturbation using Eq.~\eqref{eq:Delta_zeta}, we need to know the cosmological evolution of the homogeneous background to determine $r_{\rm d}$, $k_{\rm d}$, and $k_{\rm EMD}$ (defined in and below Eq.~\eqref{eq:Rd}).
For this purpose, it is useful to express the number of $e$-foldings $N(k)\equiv \ln(a_{\rm end}/a_k)$ from the time a $k$-mode exits the horizon, when the factor is $a_k$, to the end of inflation, when the scale factor is $a_{\rm end}$.
In a similar fashion as in~\cite{Liddle:2003as, Dodelson:2003vq}, we can write
\es{}{
{k\over a_0 H_0} = e^{-N(k)}{\sqrt{\pi} T_0 \over \sqrt{3}\cdot30^{1/4}H_0} {g_{*,s,0}^{1/3}g_{*,\rm d}^{1/4} \over g_{*,s,\rm d}^{1/3}}\left(V_k^{1/2} \over \rho_{\rm end}^{1/4}\mpl\right) \left(\rho_{\rm end} \over \rho_{\rm RH}\right)^{3w_{\rm RH}-1 \over 12(1+w_{\rm RH})}\left({\rho_{\rm d}\over \rho_{\rm EMD}}\right)^{1/12}.
}
We account for a period of matter domination after the Universe is reheated following inflation.
The energy densities $\rho_{\rm end}$, $\rho_{\rm RH}$, $\rho_{\rm EMD}$, and $\rho_{\rm d}$ denote the respective values at the end of inflation, at the end of the first reheating, at the onset of EMD, and at the end of second reheating due to $\chi$ decay. 
Here $T_0$ and $H_0$ are the present-day photon temperature and the Hubble parameter, respectively; $g_*$ and $g_{*,s}$ are respectively the effective number of degrees of freedom associated with energy and entropy density evaluated at appropriate epochs; $1/(a_0 H_0)$ is the present-day size of the comoving horizon; $V_k$ is the energy density when the $k$-mode exits the horizon; $w_{\rm RH}$ denotes the equation state of the Universe between the end of inflation and the end of the first reheating.
The above expression can be rewritten as,
\es{eq:efold}{
N(k) \approx 66.9 - \ln\left({k\over a_0H_0}\right) + \ln\left({g_{*,\rm d}^{1/4}\over g_{*,s,\rm d}^{1/3}}\right) + \ln\left(V_k^{1/2} \over \rho_{\rm end}^{1/4} \mpl\right) \\ + {1-3w_{\rm RH}\over 12(1+w_{\rm RH})}\ln\left({\rho_{\rm RH}\over \rho_{\rm end}}\right)+{1\over 12}\ln\left({\rho_{\rm d}\over \rho_{\rm EMD}}\right).
}
The quantity $w_{\rm RH}$ is model dependent~\cite{Podolsky:2005bw, Munoz:2014eqa, Lozanov:2016hid, Maity:2018qhi, Antusch:2020iyq, Allahverdi:2010xz}; however, its precise value does not impact our results in a significant manner, as explored in~\cite{Ebadi:2023xhq}.
We therefore consider the Universe to be matter-dominated during this epoch and take $w_{\rm RH}\approx 0$~\cite{Abbott:1982hn, Dolgov:1982th, Albrecht:1982mp} in the following.
We also set $k=a_0 H_0$ and $g_{*,\rm d} \approx g_{*,s,\rm d} \approx 106.75$, the value corresponding to high-temperature SM bath.
This gives,
\es{eq:efold_simple}{
N(k) \approx 66.5 + \ln\left(V_k^{1/2} \over \rho_{\rm end}^{1/4} \mpl\right) + {1\over 12}\ln\left({\rho_{\rm RH}\over \rho_{\rm end}}\right)+{1\over 12}\ln\left({\rho_{\rm d}\over \rho_{\rm EMD}}\right).
}
In the examples below, we use Eq.~\eqref{eq:efold_simple} for various choices of energy densities.
We determine $k_{\rm EMD}$ analytically assuming instantaneous transition between different regimes, obtaining
\es{eq:kemd}{
k_{\rm EMD} \approx {\rho_{\chi,\rm end} \over \rho_{\rm end}} \left(\rho_{\rm RH} \over \rho_{\rm end}\right)^{1/6} k_{\rm end}~({\rm for}~ m\sim H_{\rm end}),
}
where $\rho_{\chi,\rm end}$ is the energy density in the $\chi$ field at the end of inflation.
Here we have assumed that $\chi$ starts to oscillate soon after the end of inflation, as will be the case for Scenarios 1 (Sec.~\ref{sec:s1}) and 2 (Sec.~\ref{sec:s2}) where $m$ (mass of $\chi$) is close to $H$ during inflation.
For Scenario 3 (Sec.~\ref{sec:s3}), where $m$ can be much smaller, the corresponding expression is given in Eq.~\eqref{eq:kemd3}.
We also determine $k_{\rm d}$, the scale corresponding to $\chi$-decay,
\es{eq:kd}{
k_{\rm d} \approx \left(\tilde{\rho}_{\rm d} \over \rho_{\rm EMD}\right)^{1/6} k_{\rm EMD}.
}
Here $\tilde{\rho}_{\rm d}$ is the total energy density at the time of $\chi$-decay which we approximate as the time when $\rho_\chi/\rho_r$ reaches a maximum.
We use the same ratio to determine $r_{\rm d}$ as per Eq.~\eqref{eq:Rd}.
We illustrate the time evolution of the homogeneous energy densities in Fig.~\ref{fig:bg_evol}, obtained by solving Eq.~\eqref{eq:bg_en_density}, derived later.

\subsection{Scenario 1}
We first evaluate the average energy density in the stochastic fluctuations of the curvaton field~\cite{Ebadi:2023xhq},
\es{}{
\langle V(\chi)\rangle = {3H^4 \over 32\pi^2}\left(1 - 4 \alpha + 4\alpha{K_{3/4}(\alpha) \over K_{1/4}(\alpha)}\right)
}
where $\alpha = m^4\pi^2/(3H^4\lambda)$ and $K_n(x)$ is the modified Bessel function of the second kind.

\begin{figure}[t]
\centering
\includegraphics[width=0.75\textwidth]{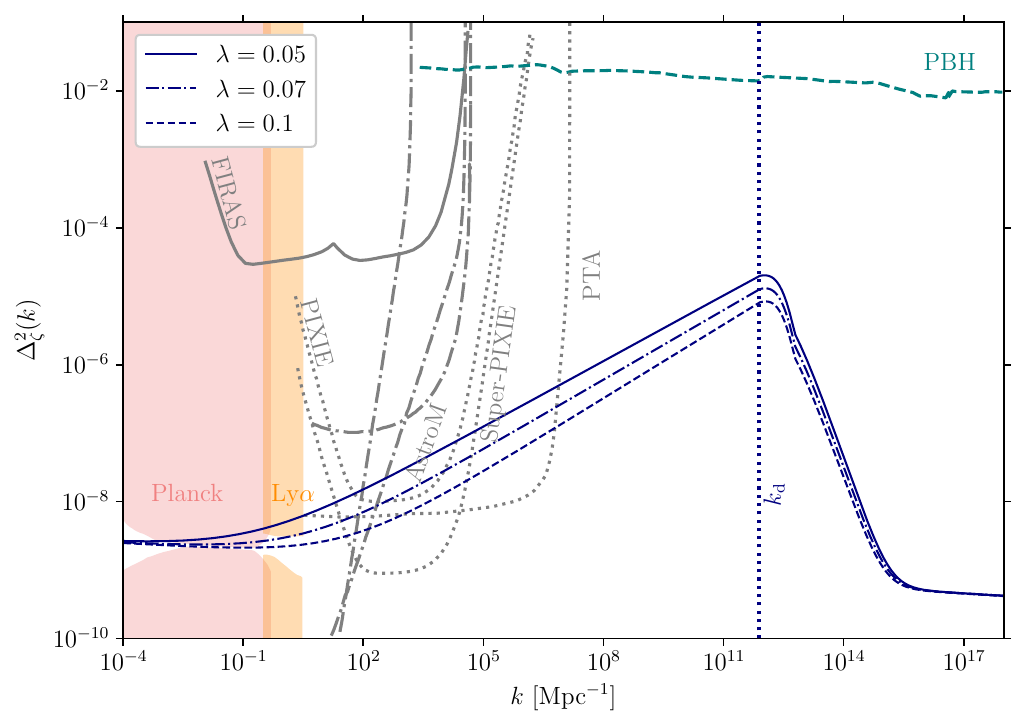}
\caption{Curvature perturbation for $m^2=0.2 H^2$ and different choices of $\lambda$ for Scenario 1. Only $\lambda=0.1$ is consistent with Planck and Ly$\alpha$ constraints~\cite{Planck:2018jri, Bird:2010mp, Chabanier:2019eai}. We show constraints from FIRAS, and projected reaches of PIXIE and Super-PIXIE~\cite{Fixsen:1996nj, 1994ApJ...420..439M, Chluba:2012gq, Chluba:2012we, Cyr:2023pgw}, along with projections from Astrometry~\cite{VanTilburg:2018ykj}, direct PTA observations~\cite{Lee:2020wfn}.    Non-observation of PBH also imposes a restriction~\cite{Carr:2020gox}. See text for the values of other parameters.}
\label{fig:stochastic_1}
\end{figure}

\paragraph{Benchmark.}
For $m^2=0.2 H^2$ and $\lambda=0.05$, $\langle V(\chi)\rangle \approx 0.02 H^4$.
We also fix $H=4\times 10^{13}$~GeV, slightly below the current upper limit~\cite{BICEP:2021xfz} and target of future B-mode experiments, $\rho_{\rm end} \simeq V_k/100$ (see, e.g.,~\cite{Abbott:1982hn, Dolgov:1982th, Albrecht:1982mp}), and a reheat temperature after inflation $T_{\rm RH}=10^{15}$~GeV.
With our choice of $\rho_{\rm d}/\rho_{\rm EMD} \approx 1.7\times 10^{-9}$, $\tilde{\rho}_{\rm d}/\rho_{\rm EMD} \approx 4.3\times 10^{-6}$, and $\rho_\chi(t_{\rm d})/\rho_r(t_{\rm d}) \approx 29$, corresponding to Fig.~\ref{fig:bg_evol}, we get
\begin{center}
\begin{tabular}{| c | c | c | c | }
	\hline
	$N$ & $k_{\rm end}~[{\rm Mpc}^{-1}]$ & $k_{\rm EMD}~[{\rm Mpc}^{-1}]$ & $k_{\rm d}~[{\rm Mpc}^{-1}]$ \\
	\hline
	$60.6$ & $4.6 \times 10^{22}$ & $6\times 10^{12}$ & $7.6\times 10^{11}$\\
	\hline
\end{tabular}
\end{center}
The result is shown in Fig.~\ref{fig:stochastic_1} for several choices of $\lambda$.
The peak of the spectrum for $\lambda = 0.1$ is at $\Delta_\zeta^2 \approx 9\times 10^{-6}$.
On the other hand, for $\lambda=0.05$ and 0.07 the curvature power spectrum is in tension with the Planck constraint.
\subsection{Scenario 2}
\begin{figure}[t]
\centering
\includegraphics[width=0.75\textwidth]{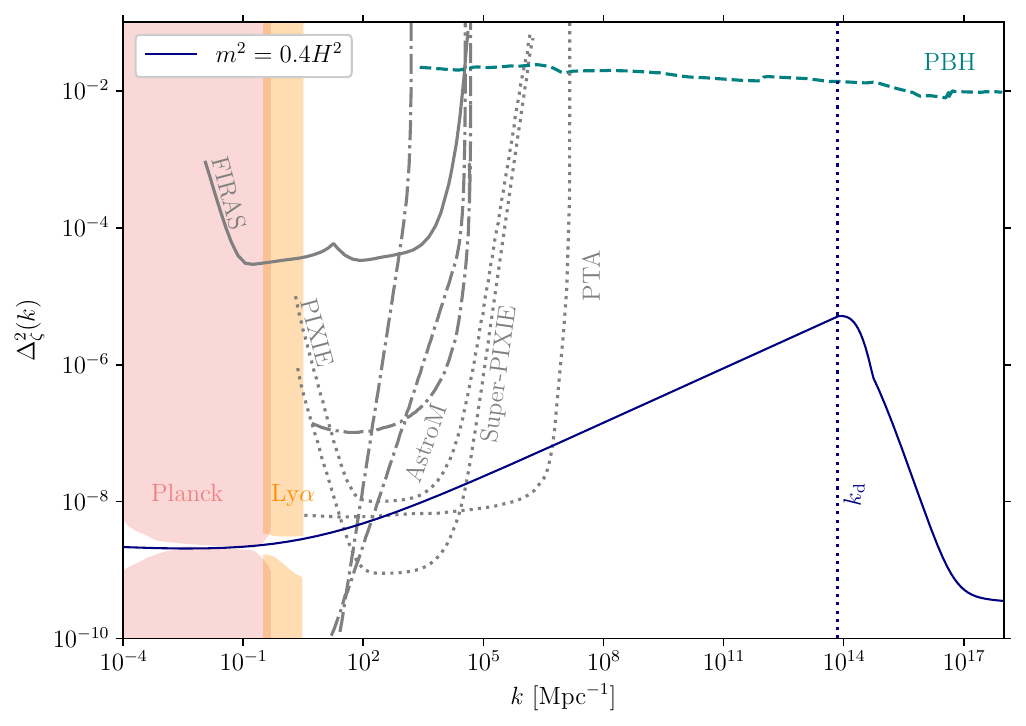}
\caption{Curvature perturbation for $m^2=0.4 H^2$ for Scenario 2. The various constraints and projections are the same as in Fig.~\ref{fig:stochastic_1}. See text for the values of other parameters.}
\label{fig:mis_1}
\end{figure}

Combining Eqs.~\eqref{eq:chi0_rat} and~\eqref{eq:mis_curv}, we determine the isocurvature power spectrum
\es{}{
\Delta_{S_\chi}^2(k) = {H^2 \over \pi^2 \tilde{\chi}_{0, \rm end}^2}\left({k \over k_{\rm end}}\right)^{2m^2/(3H^2)},
}
with $\tilde{\chi}_{0, \rm end}$ is the curvaton field value at the end of inflation.
The energy density in the curvaton field at the end of inflation is given by $\rho_{\chi, \rm end} \approx (1/2)m^2 \tilde{\chi}_{0, \rm end}^2$.
Since this can be much larger than $H^4$, both $k_{\rm EMD}$ (Eq.~\eqref{eq:kemd}) and $k_{\rm d}$ (Eq.~\eqref{eq:kemd}) can be much larger than the previous case.
\paragraph{Benchmark.}
For this we consider the same values of $H$, $\rho_{\rm end}$, $T_{\rm RH}$, $\rho_{\rm d}$, and $\tilde{\rho}_{\rm d}$ as the previous scenario.
Then for $m^2 = 0.4 H^2$ and $\tilde{\chi}_{0, \rm end} = 3 H$, we get
\begin{center}
\begin{tabular}{| c | c | c | c | }
	\hline
	$N$ & $k_{\rm end}~[{\rm Mpc}^{-1}]$ & $k_{\rm EMD}~[{\rm Mpc}^{-1}]$ & $k_{\rm d}~[{\rm Mpc}^{-1}]$ \\
	\hline
	$60.6$ & $4.6 \times 10^{22}$ & $5.4 \times 10^{14}$ & $6.9 \times 10^{13}$\\
	\hline
\end{tabular}
\end{center}
We show the resulting spectrum in Fig.~\ref{fig:mis_1}.
The shape is the same as Fig.~\ref{fig:stochastic_1}, while the location of the peak changes and the peak magnitude is now given by $\Delta_\zeta^2 \approx 5 \times 10^{-6}$.

\subsection{Scenario 3}\label{sec:s3_calc}
As explained in Sec.~\ref{sec:s3}, in this scenario we expect an almost scale-invariant spectrum of $\Delta_{S_{\chi}}^2$ for $k > k_c$; the benchmark example associated with Fig.~\ref{fig:s_dyna} implies $k_c \approx 3.2\times 10^6~{\rm Mpc}^{-1}$ for a choice of $k_*=20~{\rm Mpc}^{-1}$. 
For $k<k_c$, the spectrum would be sensitive to the precise dynamics of the radial mode and is model-dependent.
We can obtain an approximate form of the power spectrum by considering the envelope of the oscillations as shown in Fig.~\ref{fig:s_dyna}.
This gives
\es{eq:roll_approx}{
\Delta_{S_\chi}^2(k) = {H^2 \over \pi^2 (\chi_{0, \rm end} + \chi_{0,*}(k/k_*)^{\nu-3/2})^2},
} 
where $\nu = \sqrt{9/4-m_s^2/H^2}$.
Here $\chi_{0,\rm end}\equiv f_a \theta_i$ denotes the field value of the normalized Goldstone field when the radial mode has settled to its minimum. We choose the `misalignment angle' $\theta_i=1$ and therefore $\chi_{0,\rm end}$ can be read off from Fig.~\ref{fig:s_dyna}. The variable $\chi_{0,*}=s(t_*)$ where $s(t_*)$ is the value of the radial mode when the mode $k_*$ exits the horizon. Given our discussion below Eq.~\eqref{eq:iso_power_s3}, for
$k>k_c$ this asymptotes to
\es{}{
\Delta_{S_\chi}^2(k) = {H^2 \over \pi^2 \chi_{0, \rm end}^2}.
}
This is a good approximation of the actual power spectrum since the radial mode has settled to its minimum when modes with $k>k_c$ exit the horizon.

The curvature spectrum is not scale invariant since the ratio of curvaton energy density to the radiation density varies with time as per Eq.~\eqref{eq:Delta_zeta}.
In this scenario, the curvaton does not start oscillating immediately after the end of inflation since curvaton mass $m \ll H$.
Therefore, we need to modify Eq.~\eqref{eq:kemd}.
Assuming instantaneous reheating after inflation, i.e., $\rho_{\rm end} \approx \rho_{\rm RH}$, we can derive
\es{eq:kemd3}{
k_{\rm EMD} \approx {\rho_{\chi, \rm end} \over \rho_{\rm end}}\left({H_{\rm end} \over m}\right)^{3/2}k_{\rm end}~({\rm for}~\rho_{\rm end} \approx \rho_{\rm RH}).
}
The scale $k_{\rm d}$ is still given by Eq.~\eqref{eq:kd}.
With this, we can now evaluate the curvature power spectrum.

\begin{figure}[t]
\centering
\includegraphics[width=0.75\textwidth]{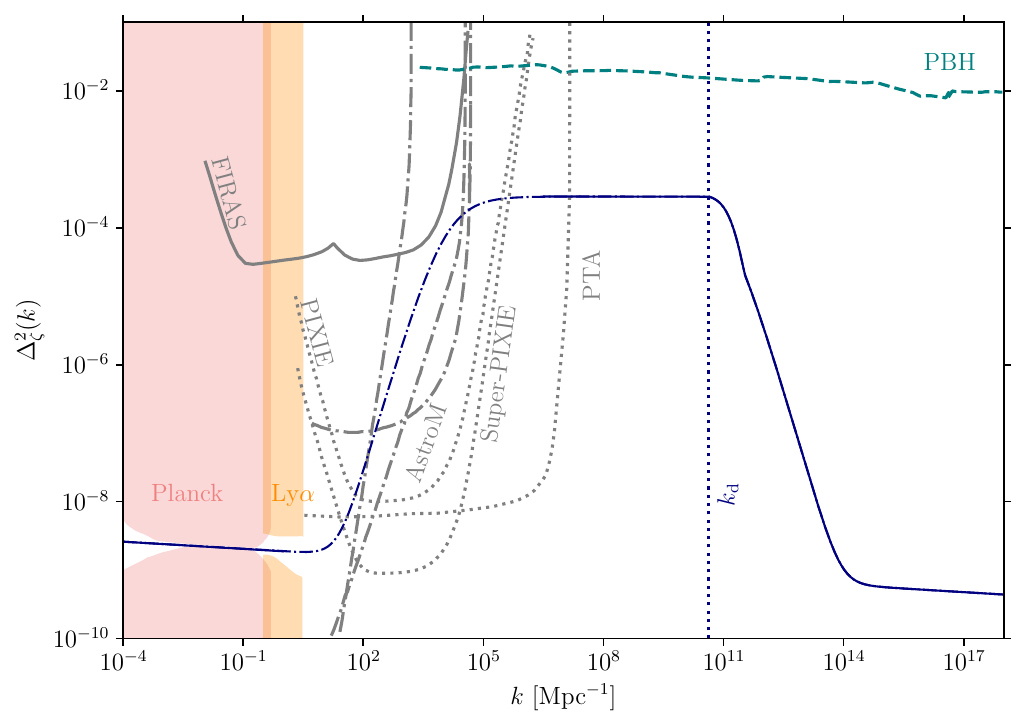}
\caption{Curvature perturbation for Scenario 3, shown in blue, for the left panel of Fig.~\ref{fig:s_dyna}. The dot-dashed portion of the curve shows an approximate modeling of the power spectrum based on Eq.~\eqref{eq:roll_approx} and is model dependent. On the other hand, for $k>k_c \approx 3.2\times 10^{6}~{\rm Mpc}^{-1}$, the spectrum can be robustly determined since the radial mode has settled to its minimum, and therefore, we show it via the solid line. See the text for the values of various parameters. The constraints and projections are the same as Fig.~\ref{fig:stochastic_1}.}
\label{fig:roll_1}
\end{figure}

\paragraph{Benchmark (a).} We choose $H=5\times 10^{12}$~GeV during inflation, which implies $T_{\rm RH} \approx 5 \times 10^{14}$~GeV, along with $m = 0.05 H$, $\chi_{0, \rm end}=6 H$, and $\lambda_\Phi = 5\times 10^{-3}$.
This implies 
\begin{center}
\begin{tabular}{| c | c | c | c | }
	\hline
	$N$ & $k_{\rm end}~[{\rm Mpc}^{-1}]$ & $k_{\rm EMD}~[{\rm Mpc}^{-1}]$ & $k_{\rm d}~[{\rm Mpc}^{-1}]$ \\
	\hline
	$59.6$ & $1.8\times 10^{22}$ & $3.3 \times 10^{11}$ & $4.2\times 10^{10}$\\
	\hline
\end{tabular}
\end{center}
The resulting spectrum is shown in Fig.~\ref{fig:roll_1}. This corresponds to the left panel of Fig.~\ref{fig:s_dyna}.

\begin{figure}[t]
\centering
\includegraphics[width=0.75\textwidth]{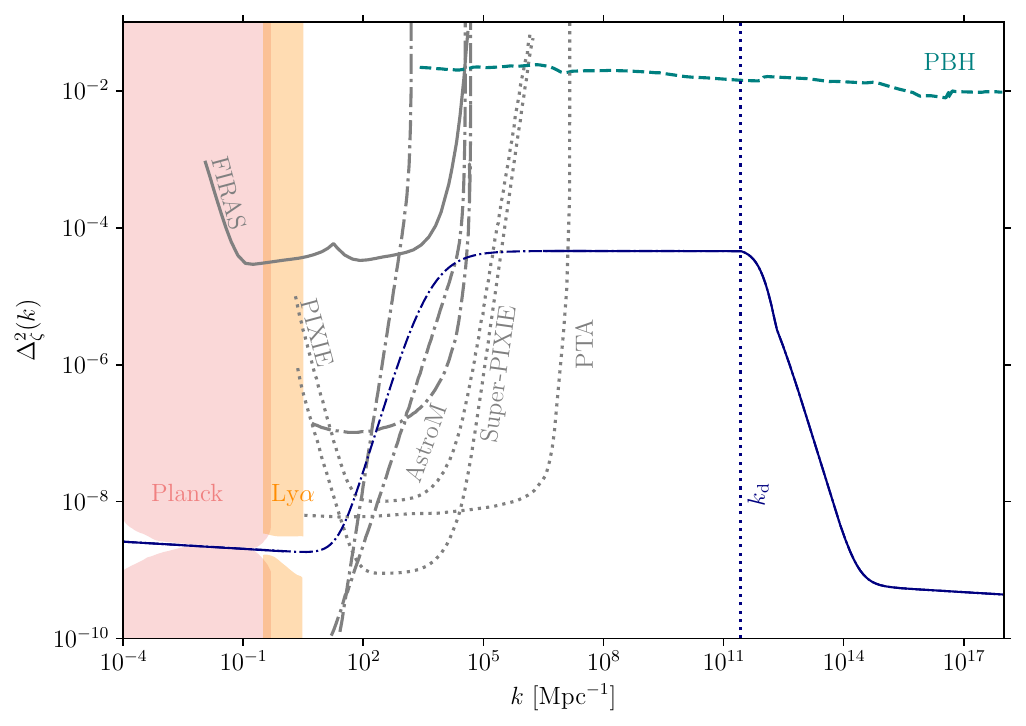}
\caption{Curvature perturbation for Scenario 3, shown in blue, for the right panel of Fig.~\ref{fig:s_dyna}. The rest of the figure follows the same convention as Fig.~\ref{fig:roll_1}.}
\label{fig:roll_2}
\end{figure}

\paragraph{Benchmark (b).}
We describe another benchmark with all the parameters identical to the above, except $\chi_{0, \rm end}=15 H$, and $\lambda_\Phi = 10^{-3}$.
This implies 
\begin{center}
\begin{tabular}{| c | c | c | c | }
	\hline
	$N$ & $k_{\rm end}~[{\rm Mpc}^{-1}]$ & $k_{\rm EMD}~[{\rm Mpc}^{-1}]$ & $k_{\rm d}~[{\rm Mpc}^{-1}]$ \\
	\hline
	$59.6$ & $1.8\times 10^{22}$ & $2.1 \times 10^{12}$ & $2.7\times 10^{11}$\\
	\hline
\end{tabular}
\end{center}
The result is shown in Fig.~\ref{fig:roll_2}.
This corresponds to the right panel of Fig.~\ref{fig:s_dyna}.

\section{Computation of Gravitational Wave Background}\label{sec:comp}
Having determined the spectrum of curvature perturbation, we now determine how it sources the SGWB.
The derivation is somewhat lengthy and hence we summarize the main results here, with a full derivation in Appendix~\ref{sec:details}.
We solve the Boltzmann equation to determine the first-order perturbations in matter, radiation, and metric, with appropriate initial conditions.
These perturbations can then be used to determine the source term for the equation of motion for the tensor fluctuations.
Upon solving that, we can determine the power spectrum for the tensor modes and determine the abundance of SGWB.  

The background equations of motion are given by,
\es{}{
3{\cal H}^2 &= 8\pi G a^2 (\bar{\rho}_m+\bar{\rho}_r),\\
{\cal H}^2 + 2{\cal H'} &= -8\pi G a^2 \bar{P}.
}
Here ${\cal H}$ is the conformal Hubble scale, $\bar{\rho}_m$ and $\bar{\rho}_r$ are homogeneous energy densities in matter and radiation, respectively.
These evolve according to
\es{}{
\bar{\rho}_r'+4{\cal H}\bar{\rho}_r &= \Gamma \bar{\rho}_m a,\\
\bar{\rho}_m'+3{\cal H}\bar{\rho}_m &= -\Gamma\bar{\rho}_m a,
}
which describes the decay of matter into radiation with $\Gamma$ providing the decay rate.
These two sets of equations fix the background evolution, subject to initial conditions that we derive in Appendix~\ref{sec:details}.

The first-order perturbations in the radiation, matter, and the metric satisfy the equations
\es{}{
\delta_m'+\Gamma a \Phi-3\Phi'+\theta_m=0,\\
\theta_m'+{\cal H}\theta_m-k^2\Phi=0,\\
\delta_r' -4\Phi' +{4\over3}\theta_r -\Gamma a {\bar{\rho}_m \over \bar{\rho}_r}(\delta_m -\delta_r+\Phi)=0,\\
\theta_r' -k^2 \Phi - {k^2\over 4}\delta_r + \Gamma a{\bar{\rho}_m \over \bar{\rho}_r}\left(\theta_r-{3\over 4}\theta_m\right)=0,\\
\Phi' +{{\cal H} \over 2}{\bar{\rho}_m \delta_m + \bar{\rho}_r \delta_r \over \bar{\rho}_m + \bar{\rho}_r}+{\cal H}\Phi+{k^2\over 3{\cal H}}\Phi = 0,
}
Here we follow the standard notation where $\delta \equiv \delta\rho/\bar{\rho}$ and $\theta$ is the velocity perturbation.
We derive the appropriate initial conditions for these perturbations in Appendix~\ref{sec:details}, tracking subdominant matter abundance at the initial time.

\begin{figure}[t]
\centering
\includegraphics[width=1\textwidth]{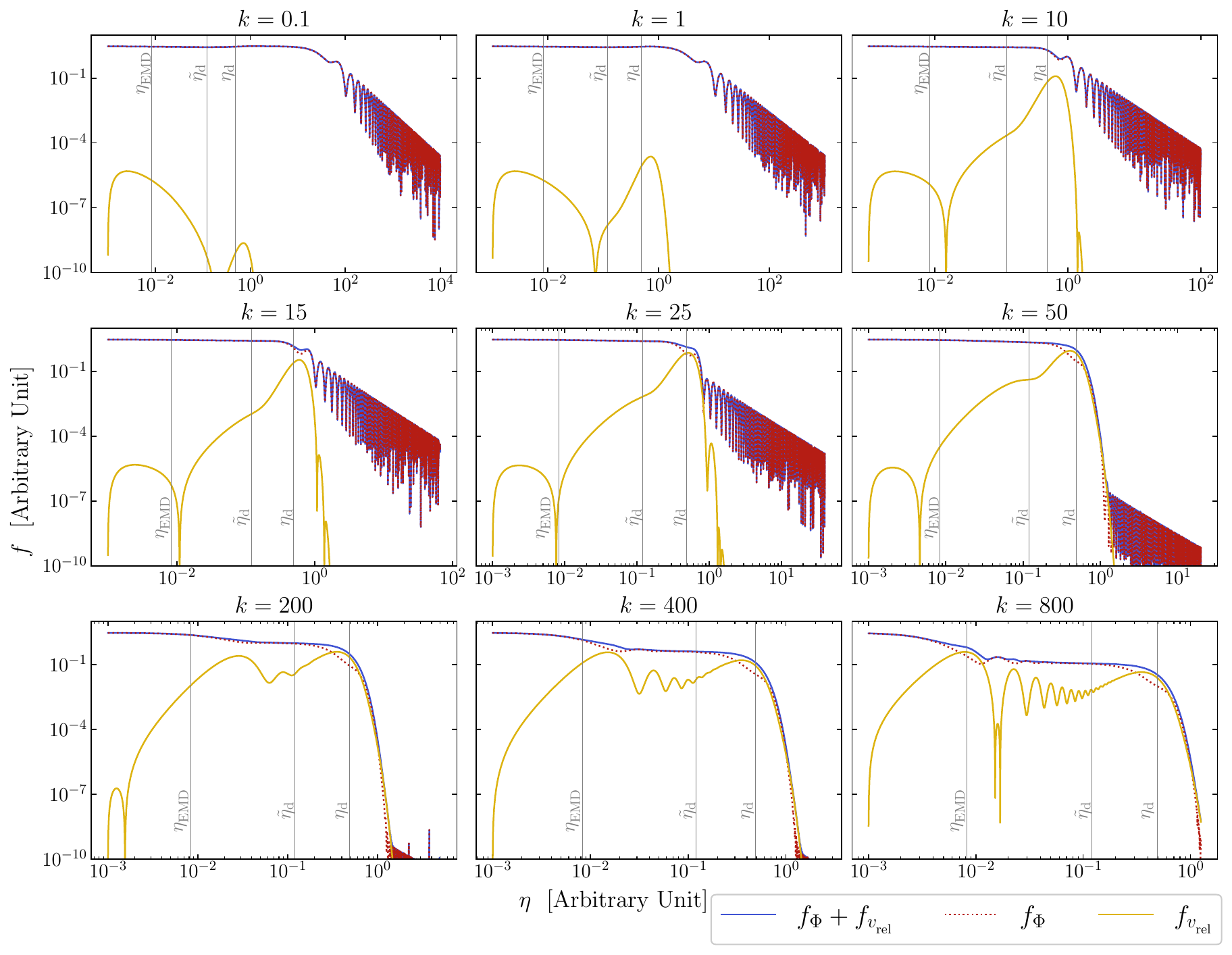}
\caption{Comparison between the contributions from Newtonian potential $f_\Phi$ and the relative velocity perturbation between matter and radiation $f_{v_{\rm rel}}$, defined in Eq.~\eqref{eq:f_def}. As seen in Eq.~\eqref{eq:fvrel}, $f_{v_{\rm rel}}$ should have a subdominant contribution when $w\approx 0$ or $w\approx 1/3$. This is reflected in the figure, as $f_{v_{\rm rel}}$ contribution decays for both $\eta < \eta_{\rm EMD}$ and $\eta > \eta_{\rm d}$, when $w\approx 1/3$. For $k\lesssim 100$, the modes reenter the horizon only after $\eta_{\rm EMD}$, hence the contribution around $\eta_{\rm EMD}$ is small. This is expected since relative velocity perturbation becomes vanishing for superhorizon modes. On the other hand, for $k\gtrsim 100$ the modes are already inside the horizon at $\eta_{\rm EMD}$. Consequently, they contribute significantly around both $\eta_{\rm EMD}$ and $\eta_{\rm d}$, when $w$ changes most significantly with time. The change in $w$ is minimal at $\tilde{\eta}_{\rm d}$, since that is deep into the EMD era with $w\approx 0$. Therefore, the contribution due to $f_{v_{\rm rel}}$ reaches an approximate minimum, as expected from Eq.~\eqref{eq:fvrel}.}
\label{fig:compare}
\end{figure}

\begin{figure}[t]
\centering
\includegraphics[width=0.7\textwidth]{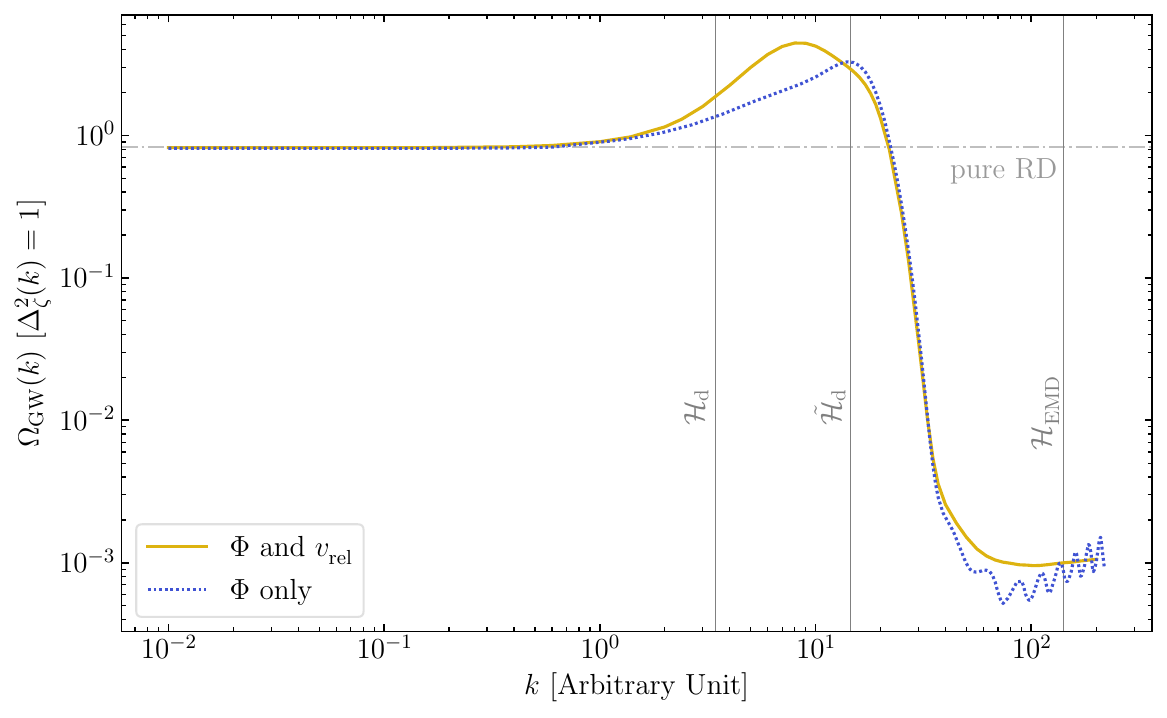}
\caption{Shape of SGWB from a scale invariant power spectrum normalized to have $\Delta_\zeta^2(k)=1$. The dot-dashed line shows the result for the pure RD era. For small values of $k\lesssim {\cal H}_{\rm d}$ that reenter the horizon after the end of EMD, the result is the same as the pure RD era, as expected. For much larger $k\gtrsim{\cal H}_{\rm EMD}$, the modes reenter the horizon prior to the onset of the EMD era. Since most of the contribution to SGWB comes from horizon crossing time, we again expect a flat spectrum, which is seen for $k\gtrsim{\cal H}_{\rm EMD}$. However, due to entropy injection from particle decay, SGWB gets diluted and $\Omega_{\rm GW}$ is smaller. For ${\cal H}_{\rm d} \lesssim k \lesssim {\cal H}_{\rm EMD}$, the effect of EMD is most prominent. This region also shows that the full contribution (solid yellow), including both the Newtonian potential and relative velocity perturbation, is larger than the contribution from Newtonian potential alone (dotted blue).}
\label{fig:scale_inv}
\end{figure}

At the second order, the graviton equation of motion is given by,
\es{eq:heom}{
h_{\vec{k}}''+2{\cal H}h_{\vec{k}}'+k^2 h_{\vec{k}} = {\cal S}_{\vec{k}},
}
where $h_{\vec{k}}$ is the Fourier transform of the metric perturbation and the source term ${\cal S}_{\vec{k}}$ is given by,
\es{}{
{\cal S}_{\vec{k}} = 4\int {\D^3 q \over (2\pi)^{3/2}}e_{ij}(\vec{k})q_i q_j\left[{12{\cal H}^2w(1-3w) \over 1+w} v_{\rm rel}(\vec{q}) v_{\rm rel}(\vec{k}-\vec{q})+{2(5+3w)\over 3(1+w)}\Phi_{\vec{q}}\Phi_{\vec{k}-\vec{q}} \right.\\ \left.+{4 \over 3(1+w)}\left(\Phi_{\vec{q}}{\Phi'_{\vec{k}-\vec{q}}\over {\cal H}}+\Phi_{\vec{k}-\vec{q}}{\Phi'_{\vec{q}}\over {\cal H}} + {\Phi'_{\vec{q}}\over {\cal H}}{\Phi'_{\vec{k}-\vec{q}}\over {\cal H}}\right)\right].
}
The expressions for the polarization tensor $e_{ij}$ is given in Appendix~\ref{sec:details}; $w=\bar{p}/ \bar{\rho}$ denotes the equation of state; and $v_{\rm rel}\equiv -(\theta_m-\theta_r)/k^2$ is the relative velocity perturbation between matter and radiation.
We study the effect of $v_{\rm rel}$ in the following.
The graviton equation of motion can be solved via the Green function,
\es{}{
G_{\vec{k}}''(\eta,\bar{\eta})+\left(k^2-{a''\over a}\right)G_{\vec{k}}(\eta,\bar{\eta})=\delta(\eta-\bar{\eta}),
}
subject to boundary conditions $G_{\vec{k}}(\bar{\eta},\bar{\eta})=0$ and $G_{\vec{k}}'(\bar{\eta},\bar{\eta})=1$.
After several intermediate steps, discussed in Appendix~\ref{sec:details},
the final result for SGWB abundance is given by,
\es{}{
\Omega_{\rm GW}(k) = \left({2\over 3}\right)^4{1\over 8 \times 24}{k^2 \over a^2 H^2}\int_{0}^\infty \D t\int_{-1}^{1} \D s \left( {t^2(2+t)^2(s^2-1)^2 \over (1+s+t)^2 (1-s+t)^2} \right.\\
\left.\overline{I(v,u,k,\eta)^2}{\Delta}_{\zeta}^2 (kv){\Delta}_{\zeta}^2 (ku)\right),
}
with
\es{}{
I(v,u,k,\eta) = k^2\int_{\eta_0}^\eta \D \bar{\eta}G_{\vec{k}}{a(\bar{\eta})\over a(\eta)}f(|\vec{q}|, |\vec{k}-\vec{q}|,\bar{\eta}),
}
and 
\es{}{
&f(|\vec{q}|, |\vec{k}-\vec{q}|,\eta)\\
=& 4 \left[{2(5+3w)\over 3(1+w)}T_{\Phi_{\vec{q}}}T_{\Phi_{\vec{k}-\vec{q}}}+{4 \over 3(1+w)}\left(T_{\Phi_{\vec{q}}}{T_{\Phi_{\vec{k}-\vec{q}}}'\over {\cal H}}+T_{\Phi_{\vec{k}-\vec{q}}}{T_{\Phi_{\vec{q}}}'\over {\cal H}} + {T_{\Phi_{\vec{q}}}'\over {\cal H}}{T_{\Phi_{\vec{k}-\vec{q}}}'\over {\cal H}}\right) \right.\\ &\quad \left.+{12{\cal H}^2w(1-3w) \over 1+w} T_{v_{\rm rel}(\vec{q})} T_{v_{\rm rel}(\vec{k}-\vec{q})}\right].
}
Here the transfer functions capture the subhorizon mode evolution, $\Phi_{\vec{k}}(\eta) = T_{\Phi_{\vec{k}}}(\eta)\tilde{\Phi}_{\vec{k}}$ and $v_{\rm rel}(\eta,\vec{k})=T_{v_{\rm rel}(\vec{k})}(\eta)\tilde{\Phi}_{\vec{k}}$ in terms of the superhorizon potential $\tilde{\Phi}_{\vec{k}} \approx (-2/3)\zeta$. 
For convenience, we can split $f$ into two contributions,
\es{eq:f_def}{
f(|\vec{q}|, |\vec{k}-\vec{q}|,\eta) = 4 (f_{\Phi}(|\vec{q}|, |\vec{k}-\vec{q}|,\eta) + f_{v_{\rm rel}}(|\vec{q}|, |\vec{k}-\vec{q}|,\eta) )
}
where 
\es{}{
&f_{\Phi}(|\vec{q}|, |\vec{k}-\vec{q}|,\eta)\\ 
=& {2(5+3w)\over 3(1+w)}T_{\Phi_{\vec{q}}}T_{\Phi_{\vec{k}-\vec{q}}}+{4 \over 3(1+w)}\left(T_{\Phi_{\vec{q}}}{T_{\Phi_{\vec{k}-\vec{q}}}'\over {\cal H}}+T_{\Phi_{\vec{k}-\vec{q}}}{T_{\Phi_{\vec{q}}}'\over {\cal H}} + {T_{\Phi_{\vec{q}}}'\over {\cal H}}{T_{\Phi_{\vec{k}-\vec{q}}}'\over {\cal H}}\right),
}
and 
\es{eq:fvrel}{
f_{v_{\rm rel}}(|\vec{q}|, |\vec{k}-\vec{q}|,\eta) = {12{\cal H}^2w(1-3w) \over 1+w} T_{v_{\rm rel}(\vec{q})} T_{v_{\rm rel}(\vec{k}-\vec{q})}.
}
While the previous literature has considered the contribution from only the $f_\Phi$ term,\footnote{See Ref.~\cite{Domenech:2020ssp} for a discussion of $f_{v_{\rm rel}}$ term in the context of a primordial black hole-dominated epoch. It was found that this term gives a subdominant contribution compared to $f_\Phi$ in that scenario.} the contribution from $f_{v_{\rm rel}}$ is important on certain length scales.
To understand this, note $f_{v_{\rm rel}}$ vanishes either when $w=0$ or $w=1/3$.
Therefore, its effect is most prominent when matter and radiation are both abundant.
Furthermore, since $v_{\rm rel}\rightarrow 0$ on superhorizon scales, the effect of $f_{v_{\rm rel}}$ is most relevant when modes are inside the horizon.
These features can be seen in Fig.~\ref{fig:compare} where we compare $f_\Phi$ and $f_{v_{\rm rel}}$.
We also track the contribution of $f_{v_{\rm rel}}$ in $\Omega_{\rm GW}$ and show the result in Fig.~\ref{fig:scale_inv}, assuming a scale invariant primordial power spectrum.
This clearly shows an enhanced signal when the contribution from $f_{v_{\rm rel}}$ is included. 
Here we have chosen the same set of parameters as in Fig.~\ref{fig:bg_evol}.
The sharp fall for larger $k$ is due to entropy dilution originating from matter decay.
We have also checked that for the same input primordial spectrum, initial conditions, and cosmological history as in~\cite{Pearce:2023kxp} and including only the $f_\Phi$ contribution, we agree with the results of~\cite{Pearce:2023kxp}.

\section{Current and Future Detectors}\label{sec:detectors}
We now show the results for the SGWB frequency spectrum using the above derivation.
We first show the spectrum for the three scenarios described above.
We then apply the Scenario 3 to the recent PTA observations~\cite{NANOGrav:2023gor, EPTA:2023sfo, EPTA:2023fyk, Reardon:2023gzh, Zic:2023gta, Xu:2023wog}, particularly focusing on NANOGrav~\cite{NANOGrav:2023gor, NANOGrav:2023hvm}.
\subsection{Predictions for the Three Scenarios}
\begin{figure}[t]
\centering
\includegraphics[width=0.7\textwidth]{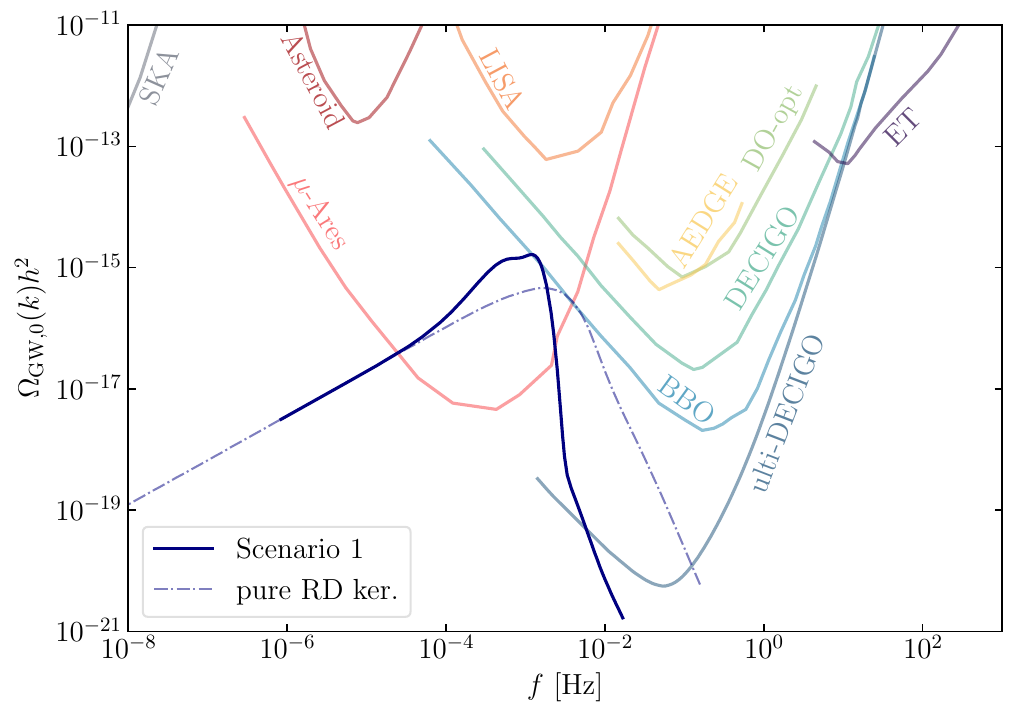}
\caption{SGWB spectrum for the Stochastic Curvaton Scenario (Scenario 1) with the curvature perturbation given in Fig.~\ref{fig:stochastic_1}. We choose $\lambda=0.1$, consistent with the CMB and Ly$\alpha$ constraints. In dot-dashed (`pure RD ker.'), we show the result for a pure RD era using the standard RD kernel to compute $\Omega_{\rm GW}$, keeping $\Delta_\zeta^2(k)$ the same. As expected based on Fig.~\ref{fig:scale_inv}, the EMD enhances the peak while suppressing the tail, due to entropy dilution from particle decay. We show the sensitivity curves for SKA, $\mu$ARES, LISA, BBO, DECIGO, DO Optimal (DO-opt), AEDGE, ET~\cite{Campeti:2020xwn}, Asteroid~\cite{Fedderke:2021kuy}, Ultimate DECIGO (ulti-DECIGO)~\cite{Braglia:2021fxn}.}
\label{fig:gw_scenario_1}
\end{figure}

\begin{figure}[t]
\centering
\includegraphics[width=0.7\textwidth]{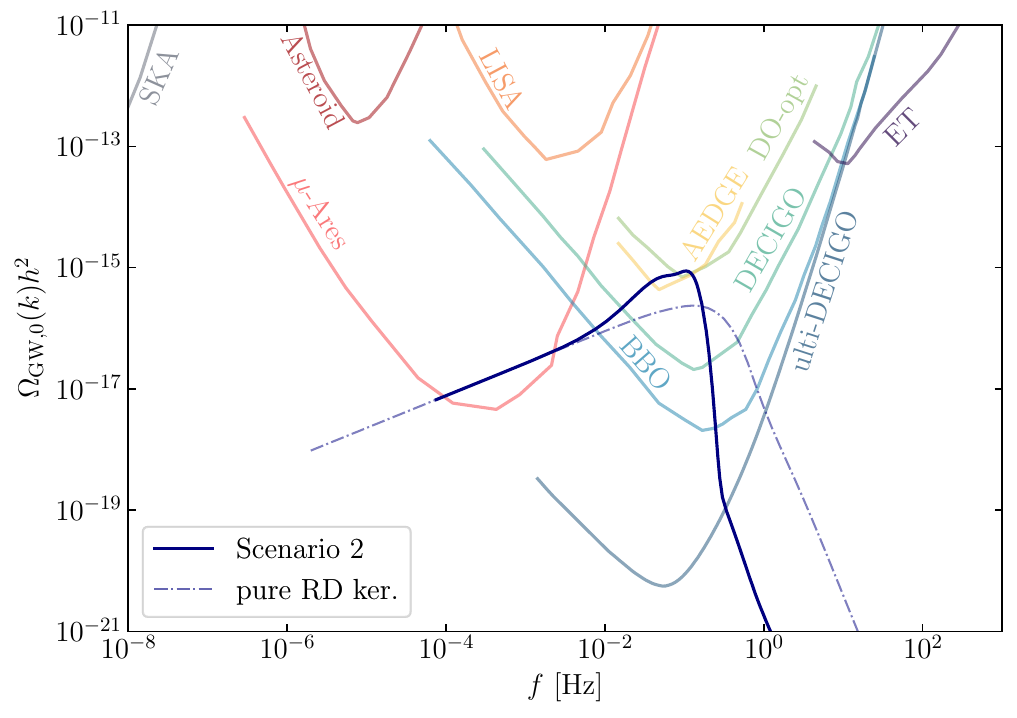}
\caption{Same as Fig.~\ref{fig:gw_scenario_1} but for the Misaligned Curvaton Scenario (Scenario 2) with the curvature perturbation given in Fig.~\ref{fig:mis_1}.}
\label{fig:gw_scenario_2}
\end{figure}

We choose the same benchmark shown in Figs.~\ref{fig:stochastic_1} and ~\ref{fig:mis_1}, and show the associated SGWB spectrum in Figs.~\ref{fig:gw_scenario_1} and~\ref{fig:gw_scenario_2}.
As explained in Sec.~\ref{sec:spectrum}, the frequency dependence is similar between Fig.~\ref{fig:gw_scenario_1} and~\ref{fig:gw_scenario_2}.
However, the spectrum for the latter is peaked at a higher frequency since the onset of the EMD happens earlier. 
Correspondingly, a different set of GW detectors, especially focused on the deci-Hz regime, become relevant. 

\begin{figure}[t]
\centering
\includegraphics[width=0.7\textwidth]{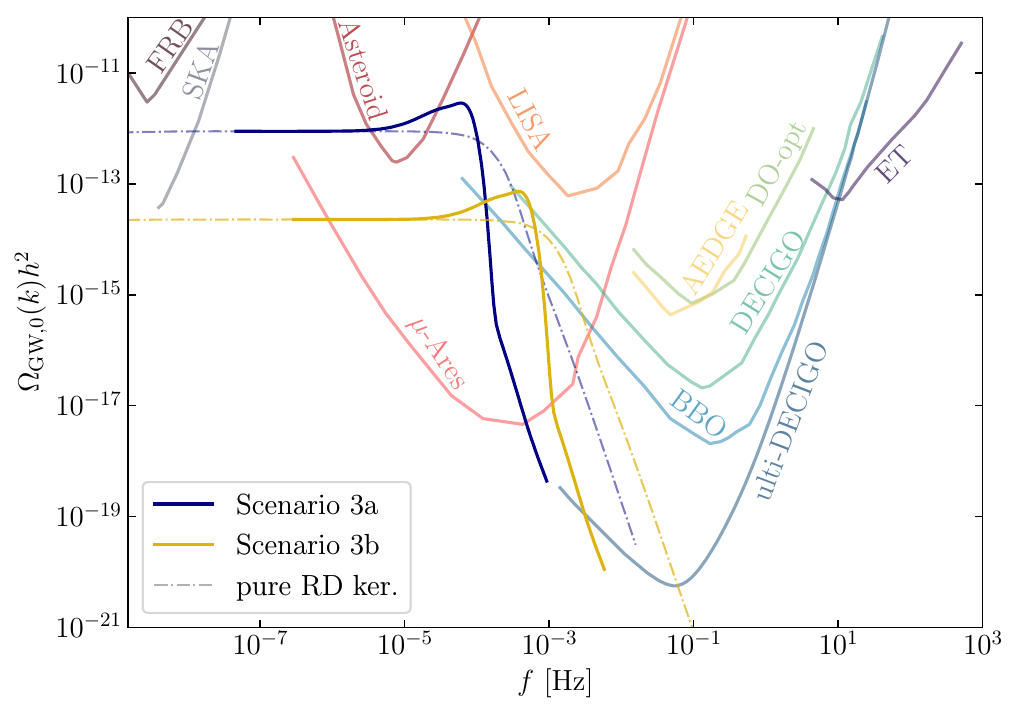}
\caption{SGWB spectrum for the Rolling Radial Mode Scenario (scenario 3). We show the results for Scenario 3a and Scenario 3b, corresponding to Figs.~\ref{fig:roll_1} and~\ref{fig:roll_2}, respectively. 
The flat low-frequency tail in this scenario could make the signal visible in multiple detectors.
The associated results for pure RD are shown via dot-dashed lines.
We also show the sensitivity that could be obtained using fast radio bursts (FRB)~\cite{Lu:2024yuo}. The rest of the projected sensitivity curves are the same as in Fig.~\ref{fig:gw_scenario_1}.}
\label{fig:gw_scenario_3}
\end{figure}

We show the results for scenario 3 for the two benchmarks discussed in Sec.~\ref{sec:spectrum}: benchmark~(a) and benchmark~(b), in Fig.~\ref{fig:gw_scenario_3}.
While the shapes are similar, benchmark~(a) gives a larger SGWB spectrum due to a smaller value of $\chi_{0,\rm end}$ which the signal is inversely proportional to.
Due to the flat spectrum for low frequencies, we typically expect to see the signal in multiple detectors, which could aid in discriminating the signal against the astrophysical foregrounds.

\subsection{Application to the Recent PTA Observations}
\begin{figure}[t]
\centering
\includegraphics[width=0.5\textwidth]{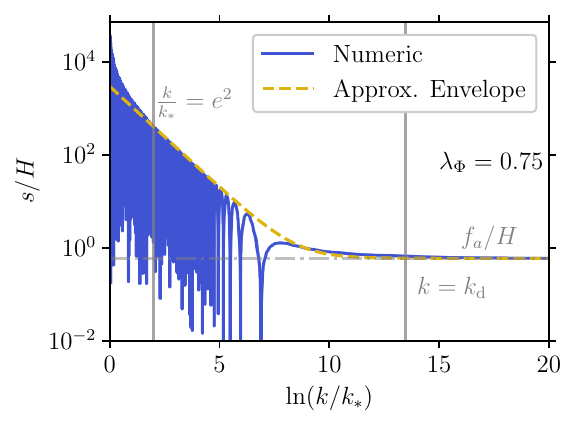}
\caption{The time evolution of the radial mode for the NANOGrav benchmark, following the same convention as Fig.~\ref{fig:s_dyna}. The initial time evolution, left edge of the plot, is dominated by the quartic coupling $\lambda_\Phi$, and hence is not well approximated by the approximate envelope. However, for $k\gtrsim e^2 k_*$, the evolution can be approximated by an effective mass $m_s$ for the radial mode.} 
\label{fig:s_dyna_nanograv}
\end{figure}

\begin{figure}[t]
\centering
\includegraphics[width=0.75\textwidth]{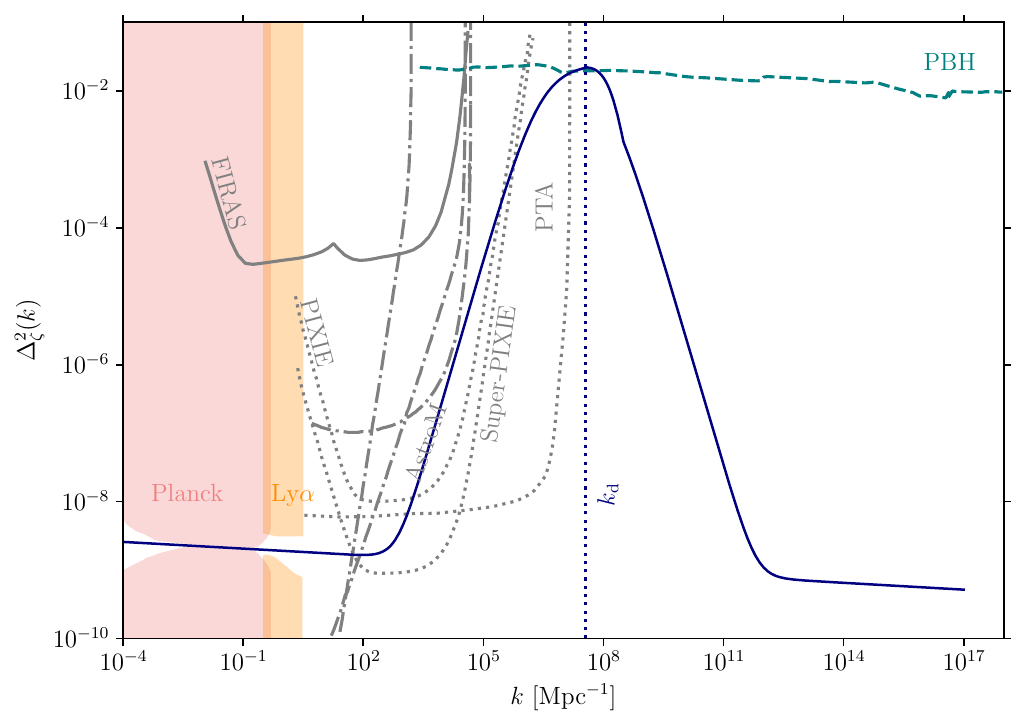}
\caption{The primordial curvature perturbation for the NANOGrav benchmark. The initial isocurvature perturbation can be obtained based on Eqs.~\eqref{eq:Delta_S} and~\eqref{eq:roll_approx}. See text for more details.}
\label{fig:Delta2_zeta_nanograv}
\end{figure}

\begin{figure}[t]
\centering
\includegraphics[width=0.7\textwidth]{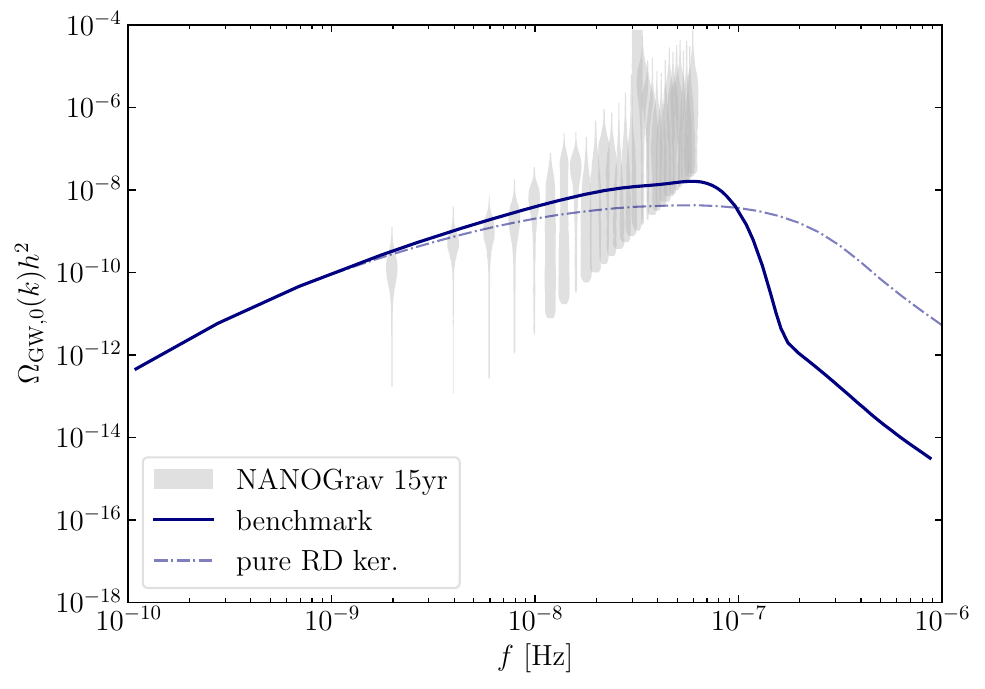}
\caption{SGWB spectrum for the NANOGrav benchmark in solid blue. The spectrum in the absence of EMD, but with the same primordial curvature perturbation, is shown in dot-dashed blue. The NANOGrav data is taken from~\cite{NANOGrav:2023hvm}.}
\label{fig:gw_nanograv}
\end{figure}

So far our discussion has been general and we have given example benchmarks for the various scenarios described above.
In this subsection, we apply Scenario~3 to the recent PTA observations, focusing on the NANOGrav data~\cite{NANOGrav:2023gor, NANOGrav:2023hvm}.
To obtain a larger strength of SGWB than considered in Fig.~\ref{fig:gw_scenario_3}, we consider a smaller value of $\chi_{0, \rm end}= 0.6 H$.
This means $\chi_0^2$ and $\langle(\delta\chi)^2\rangle$ are comparable and we evaluate $\Delta_{S_\chi}^2$ as per Eq.~\eqref{eq:Delta_S}. We also replace the constant $\chi_0$ by $\chi_{0,\rm end}+\chi_{0,*}(k/k_*)^{\nu-3/2}$, as in Eq.~\eqref{eq:roll_approx}, to capture the motion of the radial mode. 
Here we follow the same notation as Sec.~\ref{sec:s3_calc}. Namely, $\chi_{0,\rm end}$ and $\chi_{0,*}$ denote the field value of the normalized Goldstone field after the radial mode has settled into its minimum and the time when the mode $k_*$ exits the horizon, respectively. In particular, we assume the misalignment angle $\theta_i=1$, implying $\chi_{0,\rm end}=f_a=0.6H$.
We choose $H=1.9\times 10^{12}$~GeV during inflation, indicating $T_{\rm RH} \approx 3.6 \times 10^{14}$~GeV, along with $m = 0.05 H$, $\chi_{0,*} = 3.6 \times 10^4 H$, $\lambda_\Phi = 0.75$, and $k_*=50~{\rm Mpc}^{-1}$.
This implies 
\begin{center}
\begin{tabular}{| c | c | c | c | }
	\hline
	$N$ & $k_{\rm end}~[{\rm Mpc}^{-1}]$ & $k_{\rm EMD}~[{\rm Mpc}^{-1}]$ & $k_{\rm d}~[{\rm Mpc}^{-1}]$ \\
	\hline
	$59.2$ & $1.18\times 10^{22}$ & $3.14 \times 10^{8}$ & $4.0\times 10^{7}$\\
	\hline
\end{tabular}
\end{center}
The dynamics of the radial mode is shown in Fig.~\ref{fig:s_dyna_nanograv} and the primordial curvature perturbation in Fig.~\ref{fig:Delta2_zeta_nanograv}.
The final spectrum of SGWB based on this benchmark is shown in Fig.~\ref{fig:gw_nanograv}.\footnote{For this illustrative benchmark incorporating an EMD era, we do not include the effect of primordial non-Gaussianity. We leave a detailed study, which requires a dedicated numerical computation, for future work.}

\section{Conclusion and Future Directions}\label{sec:conc}
Large curvature perturbation at small length scales can source an observable stochastic gravitational wave background (SGWB), so-called `scalar-induced gravitational waves'.
In this work, we have explored several mechanisms that give rise to a curvature perturbation much larger than $\sim 10^{-5}$, a value inferred at CMB-scales.
We have shown that a curvaton field, acting as a spectator during inflation, can naturally have a blue-tilted isocurvature spectrum during inflation.
After inflation, this can source a large curvature perturbation at small scales.
In this class of scenarios, the same curvaton field can lead to a period of early matter domination (EMD).
We solved a set of Boltzmann equations that couple (decaying) matter, radiation, and scalar metric perturbation, and tracked the time evolution numerically mode-by-mode.
Using this, we computed the second-order source term for gravitational waves and obtained the strength of the SGWB.
The resulting SGWB spectrum can be observed in multiple detectors. Some of our benchmark examples particularly motivate further exploration of the `gap' between nHz and mHz frequencies which has been the focus of several recent proposals~\cite{Sesana:2019vho, Fedderke:2021kuy, Lu:2024yuo}.

While previous literature has focused on the impact of EMD on SGWB, we have taken into account several effects not considered in the previous literature.
First, we have described both the onset and end of the EMD epoch.
This has allowed us to study the full frequency dependence of SGWB due to both transitions.
Second, since the same field drives the EMD era and also generates larger curvature perturbation, we can compute full frequency dependence of SGWB without treating the small-scale curvature power spectrum as a free parameter.
Third, we have also shown for the first time that the relative velocity perturbation between (decaying) matter and radiation has an observable impact in the determination of SGWB strength.
The associated numerical framework can also be used for other cosmological scenarios with an EMD epoch, but not necessarily the same types of primordial curvature power spectrum.

For this purpose, we provide in a {\tt Mathematica} notebook~\cite{SIGW_EMD}, the kernel ${\cal K}(k,s,t)$ in Eq.~\eqref{eq:kernel_form} for a given lattice in $(s,t)$ and several example values of $k$.
This kernel only depends on the homogeneous cosmology (Fig.~\ref{fig:bg_evol}), but is independent of the primordial spectrum $\Delta_\zeta^2$.
Thus Eq.~\eqref{eq:kernel_form} can be directly used, with the provided kernels, to evaluate $\Omega_{\rm GW}$ for any choice of $\Delta_\zeta^2$.

There are several interesting future directions.
It would be interesting to study the impact of primordial non-Gaussianity on the SGWB~\cite{Adshead:2021hnm, Garcia-Saenz:2022tzu}, as an irreducible non-Gaussianity is predicted in these curvaton scenarios~\cite{Lyth:2002my, Bartolo:2003jx, Sasaki:2006kq}.
We have studied the effect of isocurvature perturbation via its conversion into curvature perturbation.
It would be useful to evaluate the time evolution of both the adiabatic and isocurvature mode by incorporating the appropriate transfer functions, see e.g.~\cite{Domenech:2020ssp, Domenech:2021and, Domenech:2023jve}.
Given the intensive numerical computations, we have focused on a single benchmark point for the background cosmology, shown in Fig.~\ref{fig:bg_evol}.
A natural next step would be to consider both shorter and longer duration of EMD in a fully flexible numerical framework (for much longer EMD, perturbations would become non-linear and some of the associated effects on SGWB have been studied in~\cite{Dalianis:2020gup, Eggemeier:2022gyo, Fernandez:2023ddy, Dalianis:2024kjr}).
We plan to address these directions in future work.

\section*{Acknowledgment}
We thank Arushi Bodas, Cara Giovanetti, Keisuke Harigaya, Keisuke Inomata, Toby Opferkuch, Ben Safdi, and Takahiro Terada for helpful discussions.
We also thank Arushi Bodas, Keisuke Harigaya, Keisuke Inomata, and Takahiro Terada for useful feedback on a draft.
SK is supported partially by the National Science Foundation (NSF) grant PHY-2210498 and the Simons Foundation.
SK thanks Aspen Center for Physics, supported by Simons Foundation (1161654, Troyer), for hospitality while this work was in progress. 
LTW is supported by the Department of Energy grant DE-SC0013642. 
We acknowledge the use of {\tt jupyter}~\cite{2016ppap.book...87K}, {\tt numpy}~\cite{harris2020array}, {\tt scipy}~\cite{2020SciPy-NMeth}, {\tt matplotlib}~\cite{Hunter:2007}, {\tt vegas}~\cite{Lepage:1977sw, Lepage:2020tgj}, {\tt jax}~\cite{jax2018github}, {\tt diffrax}~\cite{kidger2021on}, {\tt optimistix}~\cite{optimistix2024}, and {\tt Mathematica}~\cite{Mathematica}.

\appendix
\section{Details of the Computation}\label{sec:details}
We start by writing the metric fluctuations in the Newtonian gauge as,
\es{eq:newton1}{
\D s^2 = a^2(\eta)\left( -(1+2\Phi)\D \eta^2 + \left((1-2\Psi)\delta_{ij} + {1\over 2} h_{ij}\right)\D x^i \D x^j\right).
}
The energy momentum tensor is given by the sum of radiation and (decaying) matter components,
\es{}{
T_{\mu\nu} = T_{\mu\nu}^m + T_{\mu\nu}^r,
}
with $T_{\mu\nu}^m = \rho_m u^m_\mu u^m_\nu$ and $T_{\mu\nu}^r = (\rho_r+P_r) u^r_\mu u^r_\nu+P_r g_{\mu\nu}$.
The velocities satisfy the conditions $g^{\mu\nu}u^r_\mu u^r_\nu=g^{\mu\nu}u^m_\mu u^m_\nu=-1$.
To the leading order $u^r_\nu=u^m_\nu=(-a,0,0,0)$. 
The total energy-momentum tensor satisfies the equation $\nabla^\mu T_{\mu\nu}=0$ while there is an energy exchange between matter and radiation since matter decays into radiation. 
This can be parametrized as, $\nabla^\mu T^{m}_{\mu\nu}=Q_\nu$ and $\nabla^\mu T^{r}_{\mu\nu}=-Q_\nu$, with $Q_{\nu} = \Gamma T^{m\mu}_{\hspace{1em}\nu} u^m_\mu$ with $\Gamma$ the decay rate of matter.
In the following, a prime denotes a derivative with respect to conformal time $\eta$. 
\subsection{Homogeneous Equations}
From the $00$ and $ii$ components of the Einstein equations we respectively get,
\es{eq:bg}{
3{\cal H}^2 &= 8\pi G a^2 (\bar{\rho}_m+\bar{\rho}_r),\\
{\cal H}^2 + 2{\cal H'} &= -8\pi G a^2 \bar{P}.
}
From the individual energy-momentum tensors we get,
\es{eq:bg_en_density}{
\bar{\rho}_r'+4{\cal H}\bar{\rho}_r &= Q_0 = \Gamma \bar{\rho}_m a,\\
\bar{\rho}_m'+3{\cal H}\bar{\rho}_m &= -Q_0 = -\Gamma\bar{\rho}_m a,
}
which describes the decay of matter into radiation.

For $\Gamma=0$, we can analytically solve for the homogeneous background in the presence of matter and radiation. We can write the energy density as,
\es{}{
\rho = {\rho_{\rm eq} \over 2}\left((a_{\rm eq}/a)^3 + (a_{\rm eq}/a)^4)\right),
}
where $\rho_{\rm eq}$ and $a_{\rm eq}$ are the energy density and the scale factor at the time of matter-radiation equality. The Friedmann equation, following from eq.~\eqref{eq:bg} and denoting $\mpl^2 = 1/(8\pi G)$,
\es{}{
a'' = {1\over 6\mpl^2}(\rho-3P)a^3
}
implies
\es{}{
a'' = {1\over 6\mpl^2}{\rho_{\rm eq} \over 2}a_{\rm eq}^3.
}
This determines the scale factor, with the condition $a(\eta\rightarrow 0)=0$, as $a(\eta) = C_1 \eta + C_2 \eta^2$. We can rewrite this more conveniently as,
\es{eq:scale_fac}{
a = a_{\rm eq}\left((\eta/\eta_*)^2 + 2(\eta/\eta_*)\right),
}
where $a_{\rm eq}$ and $\eta_{\rm eq}$ are respectively the scale factor and conformal time at the time of matter-radiation equality and $\eta_* = \eta_{\rm eq}/(\sqrt{2}-1)$.
The expression~\eqref{eq:scale_fac} is valid when the matter energy density does not decay.
Therefore, for our scenario, the expression is approximately true much before the time of matter decay.

To obtain numerical solutions to the homogeneous background for $\Gamma\neq 0$, it is convenient to rewrite $a_{\rm eq}$ and $\eta_*$ in terms of an initial time $\eta_{\rm in}$ and the energy densities at that time $\rho_{m, \rm in}$ and $\rho_{r, \rm in}$.
This initial time is much before the epoch of matter decay and therefore the relations obtained above for $\Gamma=0$ is a good approximation.
We can use the Friedmann equations~\eqref{eq:bg}\es{eq:fried}{
a' =& {1\over \sqrt{3}\mpl}\sqrt{\rho}a^2,\\
a'' =& {1\over 6\mpl^2} \rho_m a^3,
}
and use~\eqref{eq:scale_fac} and~\eqref{eq:fried} to express,
\es{}{
\eta_* =& \eta_{\rm in} {\rhorin+\sqrt{\rhorin(\rhomin+\rhorin)} \over \rhomin},\\
a_{\rm eq} =&
{2\sqrt{3}\mpl\over \eta_{\rm in}} {\rhorin \over \rhomin} {1 \over \sqrt{\rhomin + \rhorin} + \sqrt{\rhorin}}. 
}
We use these to express the conformal Hubble rate,
\es{eq:conH}{
{\cal H} = {2\over \eta}{\eta/\eta_*+1 \over \eta/\eta_*+2},
}
in terms of the initial time and energy densities.
\subsection{First-order Perturbations}
We now study the dynamics of first-order perturbations in matter, radiation, and metric; similar derivations can be found in~\cite{Domenech:2021ztg}.
We express, $\rho_r = \bar{\rho}_r+\delta\rho_r$, $P_r = \bar{P}_r+\delta P_r$, and $\rho_m = \bar{\rho}_m+\delta\rho_m$, along with the velocities, $u_{r\mu}=(-a(1+\Phi), a \vec{v}_r)$ and $u_{m\mu}=(-a(1+\Phi), a \vec{v}_m)$.
Here $\vec{v}_m$ and $\vec{v}_r$ are first-order quantities and the temporal components have been determined from $g^{\mu\nu}u_{r\mu} u_{r\nu}=g^{\mu\nu}u_{m\mu} u_{m\nu}=-1$.

The $i\neq j$ Einstein equation fixes $\Psi=\Phi$, eliminating one of the scalar metric fluctuations.
The $00$ and the $ii$ equations give, respectively,
\es{eq:metric_fluc}{
3{\cal H}\Phi'-\triangle\Phi+8\pi G a^2(\bar{\rho}_m+\bar{\rho}_r)\Phi=3{\cal H}\Phi'-\triangle\Phi+3{\cal H}^2\Phi&=-4\pi G a^2 (\delta\rho_m +\delta\rho_r),\\
\Phi''+3{\cal H}\Phi'+2\Phi{\cal H}'+{\cal H}^2\Phi = 4\pi G a^2 \delta P_r &= 4\pi G a^2 \delta \rho_r/3,
}
where $\triangle$ denotes the 3D spatial Laplacian.
We have also used the relation $\delta P_r=\delta \rho_r/3$ for radiation.
These two equations can be combined into a single one after some algebra.
To obtain that, we write isocurvature perturbation as 
\es{eq:isocurvature_def}{
S \equiv (\delta\rho_m/\bar{\rho}_m) - (3/4)(\delta\rho_r/\bar{\rho}_r)
} and schematically write eq.~\eqref{eq:metric_fluc} as ${\cal A}=-(\delta\rho_m+\delta\rho_r)$ and ${\cal B}=\delta\rho_r/3$.
Then we can combine eq.~\eqref{eq:metric_fluc} as $c_s^2 {\cal A}+{\cal B}=-c_s^2 \bar{\rho}_m S$ which can be rewritten as,
\es{}{
\Phi''+3{\cal H}\Phi'(1+c_s^2)+(2{\cal H}'+{\cal H}^2(1+3c_s^2))\Phi-c_s^2\triangle\Phi = -4\pi G a^2 c_s^2 \bar{\rho}_m S,
}
where
\es{}{
c_s^2 = {\dot{\bar{P}}\over\dot{\bar{\rho}}} = {4\bar{\rho}_r\over(12\bar{\rho}_r+9\bar{\rho}_m)}.
}
The $0i$ Einstein equations give,
\es{}{
{\cal H}\partial_i\Phi+\partial_i\Phi'+4\pi Ga^2(\bar{\rho}_m \vec{v}_{mi}+(\bar{P}_r+\bar{\rho}_r)\vec{v}_{ri})=0.
}
Denoting $\vec{v}_r=\nabla v_r$ and $\vec{v}_m=\nabla v_m$, we can simplify the above to,
\es{eq:vel}{
{\cal H}\Phi+\Phi'+4\pi Ga^2(\bar{\rho}_m v_m+(\bar{P}_r+\bar{\rho}_r)v_r)=0.
}

Next we focus on current conservation equations.
The temporal component of $\nabla^\mu T^{m}_{\mu\nu}=Q_\nu$ gives,
\es{}{
\delta_m'+\Gamma a \Phi-3\Phi'+\triangle v_m=0,
}
while the spatial components give
\es{}{
v_m'+{\cal H}v_m+\Phi=0.
}
Here and below we follow the notation $\delta_m \equiv \delta\rho_m/\bar{\rho}_m$ and $\delta_r \equiv \delta\rho_r/\bar{\rho}_r$.
The temporal component of $\nabla^\mu T^{r}_{\mu\nu}=-Q_\nu$ gives,
\es{}{
\delta_r' = 4\Phi' -{4\over3}\nabla^2v_r +\Gamma a {\bar{\rho}_m \over \bar{\rho}_r}(\delta_m -\delta_r+\Phi),
}
while the spatial component gives,
\es{}{
v_r' + \Phi + {1\over 4}\delta_r + \Gamma a{\bar{\rho}_m \over \bar{\rho}_r}\left(v_r-{3\over 4}v_m\right)=0.
}
It is convenient to write these equations in Fourier space and we define $\theta_{m,r}=-k^2v_{m,r}$ corresponding to Fourier mode $k$.
We can then summarize the above set of equations as,
\es{eq:boltz1}{
\delta_m'+\Gamma a \Phi-3\Phi'+\theta_m=0,\\
\theta_m'+{\cal H}\theta_m-k^2\Phi=0,\\
\delta_r' -4\Phi' +{4\over3}\theta_r -\Gamma a {\bar{\rho}_m \over \bar{\rho}_r}(\delta_m -\delta_r+\Phi)=0,\\
\theta_r' -k^2 \Phi - {k^2\over 4}\delta_r + \Gamma a{\bar{\rho}_m \over \bar{\rho}_r}\left(\theta_r-{3\over 4}\theta_m\right)=0.
}
This together with the first of eq.~\eqref{eq:metric_fluc},
\es{eq:boltz2}{
\Phi' +{{\cal H} \over 2}{\bar{\rho}_m \delta_m + \bar{\rho}_r \delta_r \over \bar{\rho}_m + \bar{\rho}_r}+{\cal H}\Phi+{k^2\over 3{\cal H}}\Phi = 0,
}
constitute a closed set of first-order equations that we numerically solve.

To this end, we need the correct set of initial conditions for times much earlier than the matter decay.
Thus we can derive the initial conditions by approximating $\Gamma=0$.
For small $\eta$, we can approximate~\eqref{eq:conH},
\es{}{
{\cal H} \approx {1\over \eta}\left(1+{\eta \over 2\eta_*}\right).
}
We start with an ansatz,
\es{eq:ansatz}{
\delta_m = \sum_n a_n \eta^n,~~\delta_r = \sum_n b_n \eta^n,~~\theta_m=\sum_n c_n \eta^n,~~\theta_r=\sum_n d_n \eta^n,~~\Phi=\sum_n e_n \eta^n.
}
We have imposed the superhorizon adiabatic condition, $\delta_r/4=\delta_m/3$ which fixes $b_0 = (4/3)a_0$ and we can choose $e_0=1$ without loss of generality. It is useful to express,
\es{}{
{\bar{\rho}_m \delta_m + \bar{\rho}_r \delta_r \over \bar{\rho}_m + \bar{\rho}_r} = {{a_{\rm eq} \over a} \delta_r + \delta_m \over {a_{\rm eq} \over a}+1}.
}
We can then plug in~\eqref{eq:ansatz} in~\eqref{eq:boltz1} and \eqref{eq:boltz2} and match order-by-order in $\eta$.
The result is,
\es{eq:iclin}{
\delta_m = -(3/2) + a_1 \eta,~~\delta_r = -2 + (4a_1/3)\eta,~~\theta_m=(k^2/2)\eta,~~\theta_r=(k^2/2)\eta,~~\Phi=1+(a_1/3)\eta,
}
where $a_1=-3/(8\eta_*)$.
At this order $d_0$ is not fixed, so we {\it choose} $d_0=0$.
We can also derive the $\eta^2$ coefficients in a similar way,
\es{eq:icqua}{
a_2 = {27-28\eta_*^2k^2 \over 80\eta_*^2},~~b_2 = {27-28\eta_*^2k^2 \over 60\eta_*^2},~~c_2=-{k^2\over 8 \eta_*}=d_2,~~e_2={27-8\eta_*^2k^2 \over 240\eta_*^2}.
}
The initial conditions for pure RD can be obtained by taking the formal limit $\eta_*\rightarrow \infty$.

\subsection{Second-order Perturbations}
To derive the equation of motion obeyed by the graviton, we expand the scalar and vector metric fluctuations up to second order in the (second-order) longitudinal gauge~\cite{Acquaviva:2002ud},
\es{eq:newton2}{
\D s^2 = a^2\left( -(1+2\Phi^{(1)}+\Phi^{(2)})\D \eta^2 +\left((1-2\Psi^{(1)}-\Psi^{(2)})\delta_{ij} + {1\over 2}V_{(i,j)}+{1\over 2} h_{ij}\right)\D x^i \D x^j\right),
}
with $V_j$ divergence-free and $V_{(i,j)}\equiv \partial_i V_j+\partial_j V_i$.
To extract the traceless and transverse tensor degrees of freedom, we can use a projection tensor,
\es{}{
{\cal P}_{ij} = \delta_{ij}-{\partial_i\partial_j \over \partial^2},~~ {\cal T}_{ijkl} = {\cal P}_{ik}{\cal P}_{jl}-{1\over 2}{\cal P}_{ij}{\cal P}_{kl}.
}
We can check ${\cal T}_{ijkl}h_{kl}=h_{ij}$ with $\partial_i h_{ij}=0$ and $h_{ii}=0$ while ${\cal T}_{ijkl}V_{(k,l)}=0$.
We will not need the second-order expressions for velocities $u_{\mu}^m$ and $u_{\mu}^r$ to derive the equation of motion of the graviton $h_{ij}$.
To this end, we can evaluate the $12$ component of the Einstein equation after defining $\tilde{h}_{ij}=V_{(i,j)}+h_{ij}$,
\es{}{
&{1\over 4}\left(\tilde{h}_{12}''+2 {\cal H}\tilde{h}_{12}'-\partial_3^2 \tilde{h}_{12} + \partial_{23}\tilde{h}_{13}+8\partial_1\Phi^{(1)}\partial_2\Phi^{(1)}+\partial_{13}\tilde{h}_{23}+16\Phi^{(1)}\partial_{12}\Phi^{(1)}\right.\\
&\left. \hspace{20pt} -2\partial_{12}\Phi^{(2)}+2\partial_{12}\Psi^{(2)}-\partial_{12}\tilde{h}_{33}-2{\cal H}^2\tilde{h}_{12}-4{\cal H}'\tilde{h}_{12}\right)\\
=&4\pi G a^2\left(2\bar{\rho}_m\partial_1 v_m \partial_2v_m + {8\over 3}\bar{\rho}_r\partial_1 v_r \partial_2v_r+{1\over 3}\bar{\rho}_r\tilde{h}_{12}\right).
}
Here we use the notation $\partial_{12}\equiv \partial_1\partial_2$.
Using $\partial_i h_{ij}=0$, $h_{ii}=0$, acting by the projection tensor ${\cal T}$, and using the Friedmann equations, this can be simplified further.
Restoring generic indices $i,j$ we get:
\es{}{
h_{ij}''+2 {\cal H}h_{ij}'-\triangle h_{ij}={\cal S}_{ij},
}
where
\es{}{
{\cal S}_{ij} &= {\cal T}_{ijkl}\left(-8\partial_k \Phi^{(1)}\partial_l \Phi^{(1)}-16\Phi^{(1)}\partial_{k}\partial_{l}\Phi^{(1)}+16\pi G a^2\left(2\bar{\rho}_m\partial_k v_m \partial_l v_m + {8\over 3}\bar{\rho}_r\partial_k v_r \partial_l v_r\right)\right)\\
&= {\cal T}_{ijkl}\left(8\partial_k \Phi^{(1)}\partial_l \Phi^{(1)}+16\pi G a^2\left(2\bar{\rho}_m\partial_k v_m \partial_l v_m + {8\over 3}\bar{\rho}_r\partial_k v_r \partial_l v_r\right)\right).
}
Since the second order scalar perturbation $\Phi^{(2)}$ has dropped out, we will not include the superscript in $\Phi^{(1)}$ further.
Further simplification can be made by defining $v_{\rm rel}\equiv v_m-v_r$ and $v\equiv f_m v_m+f_r v_r$ with $f_m = \bar{\rho}_m/(\bar{\rho}_m+4\bar{\rho}_r/3)$ and $f_r = (4\bar{\rho}_r/3)/(\bar{\rho}_m+4\bar{\rho}_r/3)$, and using the Friedmann equations:
\es{}{
{\cal S}_{ij} &= {\cal T}_{ijkl}\left(8\partial_k \Phi\partial_l \Phi+32\pi G a^2\left(\bar{\rho}_m+{4\over 3}\bar{\rho}_r\right)\left(\partial_k v \partial_l v + f_m f_r\partial_k v_{\rm rel} \partial_l v_{\rm rel}\right)\right)\\
&=4{\cal T}_{ijkl}\left(2\partial_k \Phi\partial_l \Phi+{3 \bar{\rho} c_s^2 \over {\cal H}^2\bar{\rho}_r} \partial_k({\cal H}\Phi+\Phi')\partial_l({\cal H}\Phi+\Phi')+ {9{\cal H}^2\bar{\rho}_m c_s^2 \over \bar{\rho}}\partial_k v_{\rm rel} \partial_l v_{\rm rel}\right).
}

It is convenient to go to the Fourier space by writing,
\es{}{
h_{ij}(\eta, \vec{x}) = \int {\D^3\vec{k} \over (2\pi)^{3/2}} e^{i\vec{k}\cdot\vec{x}}\left(h_{\vec{k}}(\eta)e_{ij}(\vec{k}) + \bar{h}_{\vec{k}}(\eta)\bar{e}_{ij}(\vec{k})\right).
}
The polarization tensors $e_{ij}(\vec{k})$ and $\bar{e}_{ij}(\vec{k})$ are given by
\es{}{
e_{ij}(\vec{k}) \equiv {1\over \sqrt{2}} \left(e_i(\vec{k}) e_j(\vec{k}) - \bar{e}_i(\vec{k}) \bar{e}_j(\vec{k})\right),
}
and
\es{}{
\bar{e}_{ij}(\vec{k}) \equiv {1\over \sqrt{2}} \left(e_i(\vec{k}) \bar{e}_j(\vec{k}) +\bar{e}_i(\vec{k}) e_j(\vec{k})\right),
}
where $e$ and $\bar{e}$ are basis vectors orthonormal to $\vec{k}$.
The Einstein equation for $h_{\vec{k}}$ can then be derived as~\cite{Baumann:2007zm, Ananda:2006af},
\es{eq:heom}{
h_{\vec{k}}''+2{\cal H}h_{\vec{k}}'+k^2 h_{\vec{k}} = {\cal S}_{\vec{k}},
}
where 
\es{}{
{\cal S}_{\vec{k}} = 4\int {\D^3 q \over (2\pi)^{3/2}}e_{ij}(\vec{k})q_i q_j\left[{9{\cal H}^2\bar{\rho}_m c_s^2 \over \bar{\rho}} v_{\rm rel}(\vec{q}) v_{\rm rel}(\vec{k}-\vec{q})+ 2\Phi_{\vec{q}}\Phi_{\vec{k}-\vec{q}} \right.\\ \left.+{3 \bar{\rho} c_s^2 \over \bar{\rho}_r}\left(\Phi_{\vec{q}}\Phi_{\vec{k}-\vec{q}}+\Phi_{\vec{q}}{\Phi'_{\vec{k}-\vec{q}}\over {\cal H}}+\Phi_{\vec{k}-\vec{q}}{\Phi'_{\vec{q}}\over {\cal H}} + {\Phi'_{\vec{q}}\over {\cal H}}{\Phi'_{\vec{k}-\vec{q}}\over {\cal H}}\right)\right].
}
To rewrite the energy density fractions, we can introduce the equation of state of parameter $w$,
\es{}{
w \equiv {\bar{p}\over \bar{\rho}} = {\bar{\rho}_r \over 3(\bar{\rho}_r+\bar{\rho}_m)},
}
to get
\es{}{
{\cal S}_{\vec{k}} = 4\int {\D^3 q \over (2\pi)^{3/2}}e_{ij}(\vec{k})q_i q_j\left[{12{\cal H}^2w(1-3w) \over 1+w} v_{\rm rel}(\vec{q}) v_{\rm rel}(\vec{k}-\vec{q})+ {2(5+3w)\over 3(1+w)}\Phi_{\vec{q}}\Phi_{\vec{k}-\vec{q}} \right.\\ \left.+{4 \over 3(1+w)}\left(\Phi_{\vec{q}}{\Phi'_{\vec{k}-\vec{q}}\over {\cal H}}+\Phi_{\vec{k}-\vec{q}}{\Phi'_{\vec{q}}\over {\cal H}} + {\Phi'_{\vec{q}}\over {\cal H}}{\Phi'_{\vec{k}-\vec{q}}\over {\cal H}}\right)\right].
}
The factor $e_{ij}(\vec{k})q_i q_j$ can be simplified by choosing $\vec{k}$ along the $z$-axis, for example, for which $e$ and $\bar{e}$ are along $x$ and $y$ axis, respectively.
Denoting $\vec{q}\cdot \vec{k}=qk \cos(\theta)$, and $\vec{q}=(q \sin(\theta)\cos(\phi), q \sin(\theta)\sin(\phi), q\cos(\theta))$, we then get $e_{ij}q_i q_j = (q^2/\sqrt{2})\sin^2(\theta)\cos(2\phi)$. 

We can define a new variable $v_{\vec{k}}=a(\eta)h_{\vec{k}}$ in terms of which~\eqref{eq:heom} reads,
\es{}{
v''_{\vec{k}}+\left(k^2-{a''\over a}\right)v_{\vec{k}} = a {\cal S}_{\vec{k}}.
}
This can be solved using a Green function $G_{\vec{k}}(\eta,\bar{\eta})$ satisfying
\es{}{
G_{\vec{k}}''(\eta,\bar{\eta})+\left(k^2-{a''\over a}\right)G_{\vec{k}}(\eta,\bar{\eta})=\delta(\eta-\bar{\eta}),
}
with a boundary condition $G_{\vec{k}}(\bar{\eta},\bar{\eta})=0$ and $G_{\vec{k}}'(\bar{\eta},\bar{\eta})=1$, such that
\es{}{
h_{\vec{k}} = \int_{\eta_0}^\eta \D \bar{\eta} G_{\vec{k}}(\eta,\bar{\eta}) {a(\bar{\eta}) \over a(\eta)}{\cal S}_{\vec{k}}(\bar{\eta}).
}
The two point function of $h_{\vec{k}}$ is given by,
\es{}{
\langle h_{\vec{k}_1} h_{\vec{k}_2}\rangle = \int_{\eta_0}^\eta \D \bar{\eta} \int_{\eta_0}^\eta  \D \bar{\eta}' G_{\vec{k}_1}(\eta,\bar{\eta}) G_{\vec{k}_2}(\eta,\bar{\eta}'){a(\bar{\eta}) a(\bar{\eta}')\over a(\eta)^2}\langle{\cal S}_{\vec{k}_1}(\bar{\eta}){\cal S}_{\vec{k}_2}(\bar{\eta}')\rangle.
}
To compute the two point function of the source we introduce the transfer functions and write,
\es{}{
{\cal S}_{\vec{k}}(\eta) = \int {\D^3 q \over (2\pi)^{3/2}}e_{ij}(\vec{k})q_i q_j f(|\vec{q}|, |\vec{k}-\vec{q}|,\eta)\tilde{\Phi}_{\vec{q}}\tilde{\Phi}_{\vec{k}-\vec{q}},
}
where
\es{}{
f(|\vec{q}|, |\vec{k}-\vec{q}|,\eta) =& 4 \left[ {12{\cal H}^2w(1-3w) \over 1+w} T_{v_{\rm rel}(\vec{q})} T_{v_{\rm rel}(\vec{k}-\vec{q})}+{2(5+3w)\over 3(1+w)}T_{\Phi_{\vec{q}}}T_{\Phi_{\vec{k}-\vec{q}}} \right.\\ &\hspace{20pt} \left.+{4 \over 3(1+w)}\left(T_{\Phi_{\vec{q}}}{T_{\Phi_{\vec{k}-\vec{q}}}'\over {\cal H}}+T_{\Phi_{\vec{k}-\vec{q}}}{T_{\Phi_{\vec{q}}}'\over {\cal H}} + {T_{\Phi_{\vec{q}}}'\over {\cal H}}{T_{\Phi_{\vec{k}-\vec{q}}}'\over {\cal H}}\right)\right].
}
Note that $f$ is symmetric under the exchange of its first two arguments.
Here we have defined $\Phi_{\vec{k}}(\eta)=T_{\Phi_{\vec{k}}}(\eta)\tilde{\Phi}_{\vec{k}}$, and $v_{\rm rel}(\eta,\vec{k})=T_{v_{\rm rel}(\vec{k})}(\eta)\tilde{\Phi}_{\vec{k}}$, where $\tilde{\Phi}_{\vec{k}}$ is the primordial potential.
We will relate the primordial potential to curvature perturbation $\zeta$ below.
The two point function of the source then reads as,
\es{}{
\langle{\cal S}_{\vec{k}_1}(\bar{\eta}){\cal S}_{\vec{k}_2}(\bar{\eta}')\rangle = {2(2\pi^2)^2 \over (2\pi)^3}\delta^3(\vec{k}_1+\vec{k}_2)\int \D^3 q(e_{ij}q_iq_j)^2 f(|\vec{q}|, |\vec{k}_1-\vec{q}|,\bar{\eta}) \\ f(|\vec{q}|, |\vec{k}_1-\vec{q}|,\bar{\eta}'){{\Delta}_{\tilde{\Phi}}^2 (q) \over q^3} {{\Delta}_{\tilde{\Phi}}^2 (|\vec{k}_1-\vec{q}|) \over |\vec{k}_1-\vec{q}|^3},
}
for $\langle \tilde{\Phi}(\vec{k}_1) \tilde{\Phi}(\vec{k}_2)\rangle = \delta^3(\vec{k}_1+\vec{k}_2)P_{\tilde{\Phi}}(k_1)$ and $P_{\tilde{\Phi}} = (2\pi^2/k^3){\Delta}_{\tilde{\Phi}}^2$.
Here we have ignored any primordial non-Gaussianity.
In a similar way, we define $\langle h(\vec{k}_1) h(\vec{k}_2)\rangle = \delta^3(\vec{k}_1+\vec{k}_2)P_h(k_1)$ and $P_h = (2\pi^2/k^3){\Delta}_h^2$. Using the explicit expression $e_{ij}q_i q_j = (q^2/\sqrt{2})\sin^2(\theta)\cos(2\phi)$, derived above, we can perform the $\phi$ integral to get 
\es{}{
{\Delta}_h^2 = {k^3\over 4}\int_{\eta_0}^\eta \D \bar{\eta} \int_{\eta_0}^\eta  \D \bar{\eta}' G_{\vec{k}}(\eta,\bar{\eta}) G_{\vec{k}}(\eta,\bar{\eta}'){a(\bar{\eta}) a(\bar{\eta}')\over a(\eta)^2}\int \D q q^6 \sin^5\theta \D \theta f(|\vec{q}|, |\vec{k}-\vec{q}|,\bar{\eta}) \\ f(|\vec{q}|, |\vec{k}-\vec{q}|,\bar{\eta}'){{\Delta}_{\tilde{\Phi}}^2(q) \over q^3} {{\Delta}_{\tilde{\Phi}}^2(|\vec{k}-\vec{q}|) \over |\vec{k}-\vec{q}|^3}.
}
We can perform the time integrals by first defining $q=kv$, $|\vec{k}-\vec{q}|=ku$ and 
\es{}{
I(v,u,k,\eta) = k^2\int_{\eta_0}^\eta \D \bar{\eta}G_{\vec{k}}{a(\bar{\eta})\over a(\eta)}f(|\vec{q}|, |\vec{k}-\vec{q}|,\bar{\eta}).
}
Then we arrive at
\es{eq:Ph}{
{\Delta}_h^2
 = {1\over 4}\int_0^\infty \D v \int_{|1-v|}^{1+v} \D u \left(4v^2-(1-u^2+v^2)^2 \over 4vu\right)^2{\Delta}_{\tilde{\Phi}}^2 (q){\Delta}_{\tilde{\Phi}}^2 (|\vec{k}-\vec{q}|)I(v,u,k,\eta)^2.
}
The contribution from the other polarization, $\bar{h}_{\vec{k}}$, is identical; the analogous factor $\bar{e}_{ij}q_i q_j$ evaluates to $(q^2/\sqrt{2})\sin^2(\theta)\sin(2\phi)$, giving the same result after the $\phi$ integral.
The energy density in the SGWB can be computed as,
\es{}{
\Omega_{\rm GW}(k) = {1\over 3\mpl^2 H^2}{\D \rho_{\rm GW} \over \D \ln k},
}
with 
\es{}{
{\D \rho_{\rm GW} \over \D \ln k} = \mpl^2 {k^3\over 64\pi^2} \left(\langle \dot{h}(\vec{k}) \dot{h}(-\vec{k})\rangle + {k^2\over a^2}\langle h(\vec{k}) h(-\vec{k})\rangle+\langle \dot{\bar{h}}(\vec{k}) \dot{\bar{h}}(-\vec{k})\rangle + {k^2\over a^2}\langle \bar{h}(\vec{k}) \bar{h}(-\vec{k})\rangle\right).
}
Combined with the previous result we get,
\es{}{
\Omega_{\rm GW}(k) = {k^2 \over 24 a^2 H^2}{\Delta}_h^2.
}
To evaluate~\eqref{eq:Ph} numerically, it is convenient to do a further variable change:
\es{}{
u = {1\over 2}(t+s+1),~v= {1\over 2}(t-s+1),
}
using which
\es{}{
{\Delta}_h^2 = {1\over 8}\int_{0}^\infty \D t\int_{-1}^{1} \D s {t^2(2+t)^2(s^2-1)^2 \over (1+s+t)^2 (1-s+t)^2}{\Delta}_{\tilde{\Phi}}^2 (kv){\Delta}_{\tilde{\Phi}}^2 (ku)I(v,u,k,\eta)^2.
}
As a final step, we need to relate the primordial potential $\tilde{\Phi}$ to the curvature perturbation of uniform density hypersurface $\zeta$, defined as,
\es{}{
\zeta = -\Psi - {\cal H}{\delta\rho \over \bar{\rho}'}.
}
Using the first of~\eqref{eq:metric_fluc}, taking $k\rightarrow 0$ to be in the superhorizon limit, and setting $\Psi \approx \Phi$ we get
\es{}{
\zeta = -\Phi + {2\cal{H}\bar{\rho} \over \bar{\rho}'}\Phi + {2\bar{\rho} \over \bar{\rho}'}\Phi'.
}
In our computation, we start the evolution of each mode deep during the RD era ($w_{\rm ini}\rightarrow 1/3$) when superhorizon $\Phi$ is approximately constant in time.
Thus we obtain
\es{}{
\zeta \approx - {5+3w_{\rm ini} \over 3+3w_{\rm ini}}\tilde{\Phi} \rightarrow -{3\over 2}\tilde{\Phi}.
} 
Including this factor leads to our final expression,
\es{eq:omega_gw}{
\Omega_{\rm GW}(k) = \left({2\over 3}\right)^4{1\over 8 \times 24}{k^2 \over a^2 H^2}\int_{0}^\infty \D t\int_{-1}^{1} \D s \left( {t^2(2+t)^2(s^2-1)^2 \over (1+s+t)^2 (1-s+t)^2} \right.\\
 \left. {\Delta}_{\zeta}^2(kv){\Delta}_{\zeta}^2 (ku)\overline{I(v,u,k,\eta)^2} \right),
}
where the overbar on $I(v,u,k,\eta)^2$ denotes an oscillation average.

For the convenience of numerical evaluation, we take out all the terms having dependence on the final conformal time $\eta$ and define an integration kernel as
\begin{equation}
    \mathcal{K}(k,u,v,\eta) = \left( \frac{2}{3} \right)^4 \frac{1}{16} \frac{k^2}{a^2(\eta) H^2(\eta)} \overline{I(v,u,k,\eta)^2}.
\end{equation}
Here $\eta$ is a sufficiently late time after the $k$-mode has reentered the horizon and after most of the sourcing of SGWB has taken place.
With a such a choice of $\eta$ and after performing the oscillation average, it  can be checked that $\mathcal{K}(k,u,v,\eta)$ effectively becomes an $\eta$-independent function $\mathcal{K}(k,s,t)$.
We then evaluate the integral (\ref{eq:omega_gw}) as a Riemann sum
\es{eq:kernel_form}{
\Omega_{\rm GW}(k) = \frac{1}{12}  \sum_{s,t \in \Gamma(s,t)} \delta_{\Gamma(s,t)}^2 {t^2(2+t)^2(s^2-1)^2 \over (1+s+t)^2 (1-s+t)^2} \mathcal{K}(k,s,t) {\Delta}_{\zeta}^2(kv(s,t)){\Delta}_{\zeta}^2 (ku(s,t)),
}
where $\delta_{\Gamma(s,t)}$ is the lattice spacing corresponding to a defined lattice $\Gamma(s,t)$.

Given a cosmological history, such as in Fig.~\ref{fig:bg_evol}, the kernel $\mathcal{K}(k,s,t)$ is fixed.
Subsequently, Eq.~\eqref{eq:kernel_form} can be used to determine $\Omega_{\rm GW}$ for any primordial spectrum $\Delta_\zeta^2$, without a detailed and time-intensive numerical computation.
For this purpose, we provide the kernel data $\mathcal{K}(k,s,t)$ on a defined lattice $\Gamma(s,t)$ for some example values of $k$ in an accompanying {\tt Mathematica} notebook~\cite{SIGW_EMD}.

\bibliographystyle{utphys}
\bibliography{references}

\providecommand{\href}[2]{#2}\begingroup\raggedright\begin{thebibliography}{100}

\bibitem{Planck:2018jri}
{\bfseries Planck} Collaboration, Y.~Akrami {\em et~al.}, ``{Planck 2018
  results. X. Constraints on inflation},''
  \href{http://dx.doi.org/10.1051/0004-6361/201833887}{{\em Astron. Astrophys.}
  {\bfseries 641} (2020) A10},
  \href{http://arxiv.org/abs/1807.06211}{{\ttfamily arXiv:1807.06211
  [astro-ph.CO]}}.

\bibitem{Bird:2010mp}
S.~Bird, H.~V. Peiris, M.~Viel, and L.~Verde, ``{Minimally Parametric Power
  Spectrum Reconstruction from the Lyman-alpha Forest},''
  \href{http://dx.doi.org/10.1111/j.1365-2966.2011.18245.x}{{\em Mon. Not. Roy.
  Astron. Soc.} {\bfseries 413} (2011) 1717--1728},
  \href{http://arxiv.org/abs/1010.1519}{{\ttfamily arXiv:1010.1519
  [astro-ph.CO]}}.

\bibitem{Chabanier:2019eai}
S.~Chabanier, M.~Millea, and N.~Palanque-Delabrouille, ``{Matter power
  spectrum: from Ly$\alpha$ forest to CMB scales},''
  \href{http://dx.doi.org/10.1093/mnras/stz2310}{{\em Mon. Not. Roy. Astron.
  Soc.} {\bfseries 489} no.~2, (2019) 2247--2253},
  \href{http://arxiv.org/abs/1905.08103}{{\ttfamily arXiv:1905.08103
  [astro-ph.CO]}}.

\bibitem{Esteban:2023xpk}
I.~Esteban, A.~H.~G. Peter, and S.~Y. Kim, ``{Milky Way satellite velocities
  reveal the Dark Matter power spectrum at small scales},''
  \href{http://arxiv.org/abs/2306.04674}{{\ttfamily arXiv:2306.04674
  [astro-ph.CO]}}.

\bibitem{Dekker:2024nkb}
A.~Dekker and A.~Kravtsov, ``{Constraints on blue and red tilted primordial
  power spectra using dwarf galaxy properties},''
  \href{http://arxiv.org/abs/2407.04198}{{\ttfamily arXiv:2407.04198
  [astro-ph.CO]}}.

\bibitem{Gilman:2021gkj}
D.~Gilman, A.~Benson, J.~Bovy, S.~Birrer, T.~Treu, and A.~Nierenberg, ``{The
  primordial matter power spectrum on sub-galactic scales},''
  \href{http://dx.doi.org/10.1093/mnras/stac670}{{\em Mon. Not. Roy. Astron.
  Soc.} {\bfseries 512} no.~3, (2022) 3163--3188},
  \href{http://arxiv.org/abs/2112.03293}{{\ttfamily arXiv:2112.03293
  [astro-ph.CO]}}.

\bibitem{Fixsen:1996nj}
D.~J. Fixsen, E.~S. Cheng, J.~M. Gales, J.~C. Mather, R.~A. Shafer, and E.~L.
  Wright, ``{The Cosmic Microwave Background spectrum from the full COBE FIRAS
  data set},'' \href{http://dx.doi.org/10.1086/178173}{{\em Astrophys. J.}
  {\bfseries 473} (1996) 576},
  \href{http://arxiv.org/abs/astro-ph/9605054}{{\ttfamily
  arXiv:astro-ph/9605054}}.

\bibitem{1994ApJ...420..439M}
J.~C. {Mather}, E.~S. {Cheng}, D.~A. {Cottingham}, J.~{Eplee}, R.~E., D.~J.
  {Fixsen}, T.~{Hewagama}, R.~B. {Isaacman}, K.~A. {Jensen}, S.~S. {Meyer},
  P.~D. {Noerdlinger}, S.~M. {Read}, L.~P. {Rosen}, R.~A. {Shafer}, E.~L.
  {Wright}, C.~L. {Bennett}, N.~W. {Boggess}, M.~G. {Hauser}, T.~{Kelsall},
  J.~{Moseley}, S.~H., R.~F. {Silverberg}, G.~F. {Smoot}, R.~{Weiss}, and D.~T.
  {Wilkinson}, ``{Measurement of the Cosmic Microwave Background Spectrum by
  the COBE FIRAS Instrument},'' \href{http://dx.doi.org/10.1086/173574}{{\em
  ApJ} {\bfseries 420} (Jan., 1994) 439}.

\bibitem{Chluba:2012gq}
J.~Chluba, R.~Khatri, and R.~A. Sunyaev, ``{CMB at 2x2 order: The dissipation
  of primordial acoustic waves and the observable part of the associated energy
  release},'' \href{http://dx.doi.org/10.1111/j.1365-2966.2012.21474.x}{{\em
  Mon. Not. Roy. Astron. Soc.} {\bfseries 425} (2012) 1129--1169},
  \href{http://arxiv.org/abs/1202.0057}{{\ttfamily arXiv:1202.0057
  [astro-ph.CO]}}.

\bibitem{Chluba:2012we}
J.~Chluba, A.~L. Erickcek, and I.~Ben-Dayan, ``{Probing the inflaton:
  Small-scale power spectrum constraints from measurements of the CMB energy
  spectrum},'' \href{http://dx.doi.org/10.1088/0004-637X/758/2/76}{{\em
  Astrophys. J.} {\bfseries 758} (2012) 76},
  \href{http://arxiv.org/abs/1203.2681}{{\ttfamily arXiv:1203.2681
  [astro-ph.CO]}}.

\bibitem{Banik:2019cza}
N.~Banik, J.~Bovy, G.~Bertone, D.~Erkal, and T.~J.~L. de~Boer, ``{Evidence of a
  population of dark subhaloes from $Gaia$ and Pan-STARRS observations of the
  GD-1 stream},'' \href{http://dx.doi.org/10.1093/mnras/stab210}{{\em Mon. Not.
  Roy. Astron. Soc.} {\bfseries 502} no.~2, (2021) 2364--2380},
  \href{http://arxiv.org/abs/1911.02662}{{\ttfamily arXiv:1911.02662
  [astro-ph.GA]}}.

\bibitem{Ando:2022tpj}
S.~Ando, N.~Hiroshima, and K.~Ishiwata, ``{Constraining the primordial
  curvature perturbation using dark matter substructure},''
  \href{http://dx.doi.org/10.1103/PhysRevD.106.103014}{{\em Phys. Rev. D}
  {\bfseries 106} no.~10, (2022) 103014},
  \href{http://arxiv.org/abs/2207.05747}{{\ttfamily arXiv:2207.05747
  [astro-ph.CO]}}.

\bibitem{Inomata:2016uip}
K.~Inomata, M.~Kawasaki, and Y.~Tada, ``{Revisiting constraints on small scale
  perturbations from big-bang nucleosynthesis},''
  \href{http://dx.doi.org/10.1103/PhysRevD.94.043527}{{\em Phys. Rev. D}
  {\bfseries 94} no.~4, (2016) 043527},
  \href{http://arxiv.org/abs/1605.04646}{{\ttfamily arXiv:1605.04646
  [astro-ph.CO]}}.

\bibitem{Graham:2024hah}
P.~W. Graham and H.~Ramani, ``{Constraints on Dark Matter from Dynamical
  Heating of Stars in Ultrafaint Dwarfs. Part 2: Substructure and the
  Primordial Power Spectrum},''
  \href{http://arxiv.org/abs/2404.01378}{{\ttfamily arXiv:2404.01378
  [hep-ph]}}.

\bibitem{EPTA:2015qep}
{\bfseries EPTA} Collaboration, L.~Lentati {\em et~al.}, ``{European Pulsar
  Timing Array Limits On An Isotropic Stochastic Gravitational-Wave
  Background},'' \href{http://dx.doi.org/10.1093/mnras/stv1538}{{\em Mon. Not.
  Roy. Astron. Soc.} {\bfseries 453} no.~3, (2015) 2576--2598},
  \href{http://arxiv.org/abs/1504.03692}{{\ttfamily arXiv:1504.03692
  [astro-ph.CO]}}.

\bibitem{NANOGrav:2023gor}
{\bfseries NANOGrav} Collaboration, G.~Agazie {\em et~al.}, ``{The NANOGrav 15
  yr Data Set: Evidence for a Gravitational-wave Background},''
  \href{http://dx.doi.org/10.3847/2041-8213/acdac6}{{\em Astrophys. J. Lett.}
  {\bfseries 951} no.~1, (2023) L8},
  \href{http://arxiv.org/abs/2306.16213}{{\ttfamily arXiv:2306.16213
  [astro-ph.HE]}}.

\bibitem{EPTA:2023sfo}
{\bfseries EPTA} Collaboration, J.~Antoniadis {\em et~al.}, ``{The second data
  release from the European Pulsar Timing Array - I. The dataset and timing
  analysis},'' \href{http://dx.doi.org/10.1051/0004-6361/202346841}{{\em
  Astron. Astrophys.} {\bfseries 678} (2023) A48},
  \href{http://arxiv.org/abs/2306.16224}{{\ttfamily arXiv:2306.16224
  [astro-ph.HE]}}.

\bibitem{EPTA:2023fyk}
{\bfseries EPTA} Collaboration, J.~Antoniadis {\em et~al.}, ``{The second data
  release from the European Pulsar Timing Array III. Search for gravitational
  wave signals},'' \href{http://arxiv.org/abs/2306.16214}{{\ttfamily
  arXiv:2306.16214 [astro-ph.HE]}}.

\bibitem{Reardon:2023gzh}
D.~J. Reardon {\em et~al.}, ``{Search for an Isotropic Gravitational-wave
  Background with the Parkes Pulsar Timing Array},''
  \href{http://dx.doi.org/10.3847/2041-8213/acdd02}{{\em Astrophys. J. Lett.}
  {\bfseries 951} no.~1, (2023) L6},
  \href{http://arxiv.org/abs/2306.16215}{{\ttfamily arXiv:2306.16215
  [astro-ph.HE]}}.

\bibitem{Zic:2023gta}
A.~Zic {\em et~al.}, ``{The Parkes Pulsar Timing Array third data release},''
  \href{http://dx.doi.org/10.1017/pasa.2023.36}{{\em Publ. Astron. Soc.
  Austral.} {\bfseries 40} (2023) e049},
  \href{http://arxiv.org/abs/2306.16230}{{\ttfamily arXiv:2306.16230
  [astro-ph.HE]}}.

\bibitem{Xu:2023wog}
H.~Xu {\em et~al.}, ``{Searching for the Nano-Hertz Stochastic Gravitational
  Wave Background with the Chinese Pulsar Timing Array Data Release I},''
  \href{http://dx.doi.org/10.1088/1674-4527/acdfa5}{{\em Res. Astron.
  Astrophys.} {\bfseries 23} no.~7, (2023) 075024},
  \href{http://arxiv.org/abs/2306.16216}{{\ttfamily arXiv:2306.16216
  [astro-ph.HE]}}.

\bibitem{Chung:2023syw}
D.~J.~H. Chung, M.~M\"unchmeyer, and S.~C. Tadepalli, ``{Search for
  isocurvature with large-scale structure: A forecast for Euclid and MegaMapper
  using EFTofLSS},'' \href{http://dx.doi.org/10.1103/PhysRevD.108.103542}{{\em
  Phys. Rev. D} {\bfseries 108} no.~10, (2023) 103542},
  \href{http://arxiv.org/abs/2306.09456}{{\ttfamily arXiv:2306.09456
  [astro-ph.CO]}}.

\bibitem{Sekiguchi:2013lma}
T.~Sekiguchi, H.~Tashiro, J.~Silk, and N.~Sugiyama, ``{Cosmological signatures
  of tilted isocurvature perturbations: reionization and 21cm fluctuations},''
  \href{http://dx.doi.org/10.1088/1475-7516/2014/03/001}{{\em JCAP} {\bfseries
  03} (2014) 001}, \href{http://arxiv.org/abs/1311.3294}{{\ttfamily
  arXiv:1311.3294 [astro-ph.CO]}}.

\bibitem{deKruijf:2024voc}
J.~de~Kruijf, E.~Vanzan, K.~K. Boddy, A.~Raccanelli, and N.~Bartolo,
  ``{Searching for blue in the dark},''
  \href{http://arxiv.org/abs/2408.04991}{{\ttfamily arXiv:2408.04991
  [astro-ph.CO]}}.

\bibitem{VanTilburg:2018ykj}
K.~Van~Tilburg, A.-M. Taki, and N.~Weiner, ``{Halometry from Astrometry},''
  \href{http://dx.doi.org/10.1088/1475-7516/2018/07/041}{{\em JCAP} {\bfseries
  07} (2018) 041}, \href{http://arxiv.org/abs/1804.01991}{{\ttfamily
  arXiv:1804.01991 [astro-ph.CO]}}.

\bibitem{Lee:2020wfn}
V.~S.~H. Lee, A.~Mitridate, T.~Trickle, and K.~M. Zurek, ``{Probing Small-Scale
  Power Spectra with Pulsar Timing Arrays},''
  \href{http://dx.doi.org/10.1007/JHEP06(2021)028}{{\em JHEP} {\bfseries 06}
  (2021) 028}, \href{http://arxiv.org/abs/2012.09857}{{\ttfamily
  arXiv:2012.09857 [astro-ph.CO]}}.

\bibitem{Xiao:2024qay}
H.~Xiao, L.~Dai, and M.~McQuinn, ``{Detecting dark matter substructures on
  small scales with fast radio bursts},''
  \href{http://dx.doi.org/10.1103/PhysRevD.110.023516}{{\em Phys. Rev. D}
  {\bfseries 110} no.~2, (2024) 023516},
  \href{http://arxiv.org/abs/2401.08862}{{\ttfamily arXiv:2401.08862
  [astro-ph.CO]}}.

\bibitem{Ananda:2006af}
K.~N. Ananda, C.~Clarkson, and D.~Wands, ``{The Cosmological gravitational wave
  background from primordial density perturbations},''
  \href{http://dx.doi.org/10.1103/PhysRevD.75.123518}{{\em Phys. Rev. D}
  {\bfseries 75} (2007) 123518},
  \href{http://arxiv.org/abs/gr-qc/0612013}{{\ttfamily arXiv:gr-qc/0612013}}.

\bibitem{Baumann:2007zm}
D.~Baumann, P.~J. Steinhardt, K.~Takahashi, and K.~Ichiki, ``{Gravitational
  Wave Spectrum Induced by Primordial Scalar Perturbations},''
  \href{http://dx.doi.org/10.1103/PhysRevD.76.084019}{{\em Phys. Rev. D}
  {\bfseries 76} (2007) 084019},
  \href{http://arxiv.org/abs/hep-th/0703290}{{\ttfamily arXiv:hep-th/0703290}}.

\bibitem{Domenech:2021ztg}
G.~Dom\`enech, ``{Scalar Induced Gravitational Waves Review},''
  \href{http://dx.doi.org/10.3390/universe7110398}{{\em Universe} {\bfseries 7}
  no.~11, (2021) 398}, \href{http://arxiv.org/abs/2109.01398}{{\ttfamily
  arXiv:2109.01398 [gr-qc]}}.

\bibitem{Inomata:2018epa}
K.~Inomata and T.~Nakama, ``{Gravitational waves induced by scalar
  perturbations as probes of the small-scale primordial spectrum},''
  \href{http://dx.doi.org/10.1103/PhysRevD.99.043511}{{\em Phys. Rev. D}
  {\bfseries 99} no.~4, (2019) 043511},
  \href{http://arxiv.org/abs/1812.00674}{{\ttfamily arXiv:1812.00674
  [astro-ph.CO]}}.

\bibitem{Green:2020jor}
A.~M. Green and B.~J. Kavanagh, ``{Primordial Black Holes as a dark matter
  candidate},'' \href{http://dx.doi.org/10.1088/1361-6471/abc534}{{\em J. Phys.
  G} {\bfseries 48} no.~4, (2021) 043001},
  \href{http://arxiv.org/abs/2007.10722}{{\ttfamily arXiv:2007.10722
  [astro-ph.CO]}}.

\bibitem{Carr:2020xqk}
B.~Carr and F.~Kuhnel, ``{Primordial Black Holes as Dark Matter: Recent
  Developments},''
  \href{http://dx.doi.org/10.1146/annurev-nucl-050520-125911}{{\em Ann. Rev.
  Nucl. Part. Sci.} {\bfseries 70} (2020) 355--394},
  \href{http://arxiv.org/abs/2006.02838}{{\ttfamily arXiv:2006.02838
  [astro-ph.CO]}}.

\bibitem{Ozsoy:2023ryl}
O.~\"Ozsoy and G.~Tasinato, ``{Inflation and Primordial Black Holes},''
  \href{http://dx.doi.org/10.3390/universe9050203}{{\em Universe} {\bfseries 9}
  no.~5, (2023) 203}, \href{http://arxiv.org/abs/2301.03600}{{\ttfamily
  arXiv:2301.03600 [astro-ph.CO]}}.

\bibitem{Ivanov:1994pa}
P.~Ivanov, P.~Naselsky, and I.~Novikov, ``{Inflation and primordial black holes
  as dark matter},'' \href{http://dx.doi.org/10.1103/PhysRevD.50.7173}{{\em
  Phys. Rev. D} {\bfseries 50} (1994) 7173--7178}.

\bibitem{Garcia-Bellido:2017mdw}
J.~Garcia-Bellido and E.~Ruiz~Morales, ``{Primordial black holes from single
  field models of inflation},''
  \href{http://dx.doi.org/10.1016/j.dark.2017.09.007}{{\em Phys. Dark Univ.}
  {\bfseries 18} (2017) 47--54},
  \href{http://arxiv.org/abs/1702.03901}{{\ttfamily arXiv:1702.03901
  [astro-ph.CO]}}.

\bibitem{Ballesteros:2017fsr}
G.~Ballesteros and M.~Taoso, ``{Primordial black hole dark matter from single
  field inflation},'' \href{http://dx.doi.org/10.1103/PhysRevD.97.023501}{{\em
  Phys. Rev. D} {\bfseries 97} no.~2, (2018) 023501},
  \href{http://arxiv.org/abs/1709.05565}{{\ttfamily arXiv:1709.05565
  [hep-ph]}}.

\bibitem{Hertzberg:2017dkh}
M.~P. Hertzberg and M.~Yamada, ``{Primordial Black Holes from Polynomial
  Potentials in Single Field Inflation},''
  \href{http://dx.doi.org/10.1103/PhysRevD.97.083509}{{\em Phys. Rev. D}
  {\bfseries 97} no.~8, (2018) 083509},
  \href{http://arxiv.org/abs/1712.09750}{{\ttfamily arXiv:1712.09750
  [astro-ph.CO]}}.

\bibitem{Stewart:1996ey}
E.~D. Stewart, ``{Flattening the inflaton's potential with quantum
  corrections},'' \href{http://dx.doi.org/10.1016/S0370-2693(96)01458-X}{{\em
  Phys. Lett. B} {\bfseries 391} (1997) 34--38},
  \href{http://arxiv.org/abs/hep-ph/9606241}{{\ttfamily arXiv:hep-ph/9606241}}.

\bibitem{Stewart:1997wg}
E.~D. Stewart, ``{Flattening the inflaton's potential with quantum corrections.
  2.},'' \href{http://dx.doi.org/10.1103/PhysRevD.56.2019}{{\em Phys. Rev. D}
  {\bfseries 56} (1997) 2019--2023},
  \href{http://arxiv.org/abs/hep-ph/9703232}{{\ttfamily arXiv:hep-ph/9703232}}.

\bibitem{Leach:2000ea}
S.~M. Leach, I.~J. Grivell, and A.~R. Liddle, ``{Black hole constraints on the
  running mass inflation model},''
  \href{http://dx.doi.org/10.1103/PhysRevD.62.043516}{{\em Phys. Rev. D}
  {\bfseries 62} (2000) 043516},
  \href{http://arxiv.org/abs/astro-ph/0004296}{{\ttfamily
  arXiv:astro-ph/0004296}}.

\bibitem{Kohri:2007qn}
K.~Kohri, D.~H. Lyth, and A.~Melchiorri, ``{Black hole formation and slow-roll
  inflation},'' \href{http://dx.doi.org/10.1088/1475-7516/2008/04/038}{{\em
  JCAP} {\bfseries 04} (2008) 038},
  \href{http://arxiv.org/abs/0711.5006}{{\ttfamily arXiv:0711.5006 [hep-ph]}}.

\bibitem{Alabidi:2009bk}
L.~Alabidi and K.~Kohri, ``{Generating Primordial Black Holes Via Hilltop-Type
  Inflation Models},'' \href{http://dx.doi.org/10.1103/PhysRevD.80.063511}{{\em
  Phys. Rev. D} {\bfseries 80} (2009) 063511},
  \href{http://arxiv.org/abs/0906.1398}{{\ttfamily arXiv:0906.1398
  [astro-ph.CO]}}.

\bibitem{Baumann:2014nda}
D.~Baumann and L.~McAllister,
  \href{http://dx.doi.org/10.1017/CBO9781316105733}{{\em {Inflation and String
  Theory}}}.
\newblock Cambridge Monographs on Mathematical Physics. Cambridge University
  Press, 5, 2015.
\newblock \href{http://arxiv.org/abs/1404.2601}{{\ttfamily arXiv:1404.2601
  [hep-th]}}.

\bibitem{Kasuya:2009up}
S.~Kasuya and M.~Kawasaki, ``{Axion isocurvature fluctuations with extremely
  blue spectrum},'' \href{http://dx.doi.org/10.1103/PhysRevD.80.023516}{{\em
  Phys. Rev. D} {\bfseries 80} (2009) 023516},
  \href{http://arxiv.org/abs/0904.3800}{{\ttfamily arXiv:0904.3800
  [astro-ph.CO]}}.

\bibitem{Kawasaki:2012wr}
M.~Kawasaki, N.~Kitajima, and T.~T. Yanagida, ``{Primordial black hole
  formation from an axionlike curvaton model},''
  \href{http://dx.doi.org/10.1103/PhysRevD.87.063519}{{\em Phys. Rev. D}
  {\bfseries 87} no.~6, (2013) 063519},
  \href{http://arxiv.org/abs/1207.2550}{{\ttfamily arXiv:1207.2550 [hep-ph]}}.

\bibitem{Kawasaki:2013xsa}
M.~Kawasaki, N.~Kitajima, and S.~Yokoyama, ``{Gravitational waves from a
  curvaton model with blue spectrum},''
  \href{http://dx.doi.org/10.1088/1475-7516/2013/08/042}{{\em JCAP} {\bfseries
  08} (2013) 042}, \href{http://arxiv.org/abs/1305.4464}{{\ttfamily
  arXiv:1305.4464 [astro-ph.CO]}}.

\bibitem{Ando:2018nge}
K.~Ando, M.~Kawasaki, and H.~Nakatsuka, ``{Formation of primordial black holes
  in an axionlike curvaton model},''
  \href{http://dx.doi.org/10.1103/PhysRevD.98.083508}{{\em Phys. Rev. D}
  {\bfseries 98} no.~8, (2018) 083508},
  \href{http://arxiv.org/abs/1805.07757}{{\ttfamily arXiv:1805.07757
  [astro-ph.CO]}}.

\bibitem{Ebadi:2023xhq}
R.~Ebadi, S.~Kumar, A.~McCune, H.~Tai, and L.-T. Wang, ``{Gravitational waves
  from stochastic scalar fluctuations},''
  \href{http://dx.doi.org/10.1103/PhysRevD.109.083519}{{\em Phys. Rev. D}
  {\bfseries 109} no.~8, (2024) 083519},
  \href{http://arxiv.org/abs/2307.01248}{{\ttfamily arXiv:2307.01248
  [astro-ph.CO]}}.

\bibitem{Inomata:2023drn}
K.~Inomata, M.~Kawasaki, K.~Mukaida, and T.~T. Yanagida, ``{Axion curvaton
  model for the gravitational waves observed by pulsar timing arrays},''
  \href{http://dx.doi.org/10.1103/PhysRevD.109.043508}{{\em Phys. Rev. D}
  {\bfseries 109} no.~4, (2024) 043508},
  \href{http://arxiv.org/abs/2309.11398}{{\ttfamily arXiv:2309.11398
  [astro-ph.CO]}}.

\bibitem{Espinosa:2018eve}
J.~R. Espinosa, D.~Racco, and A.~Riotto, ``{A Cosmological Signature of the SM
  Higgs Instability: Gravitational Waves},''
  \href{http://dx.doi.org/10.1088/1475-7516/2018/09/012}{{\em JCAP} {\bfseries
  09} (2018) 012}, \href{http://arxiv.org/abs/1804.07732}{{\ttfamily
  arXiv:1804.07732 [hep-ph]}}.

\bibitem{Kohri:2018awv}
K.~Kohri and T.~Terada, ``{Semianalytic calculation of gravitational wave
  spectrum nonlinearly induced from primordial curvature perturbations},''
  \href{http://dx.doi.org/10.1103/PhysRevD.97.123532}{{\em Phys. Rev. D}
  {\bfseries 97} no.~12, (2018) 123532},
  \href{http://arxiv.org/abs/1804.08577}{{\ttfamily arXiv:1804.08577 [gr-qc]}}.

\bibitem{Domenech:2019quo}
G.~Dom\`enech, ``{Induced gravitational waves in a general cosmological
  background},'' \href{http://dx.doi.org/10.1142/S0218271820500285}{{\em Int.
  J. Mod. Phys. D} {\bfseries 29} no.~03, (2020) 2050028},
  \href{http://arxiv.org/abs/1912.05583}{{\ttfamily arXiv:1912.05583 [gr-qc]}}.

\bibitem{SIGW_EMD}
\url{https://github.com/soubhikk/SIGW_EMD}.

\bibitem{NANOGrav:2023hvm}
{\bfseries NANOGrav} Collaboration, A.~Afzal {\em et~al.}, ``{The NANOGrav 15
  yr Data Set: Search for Signals from New Physics},''
  \href{http://dx.doi.org/10.3847/2041-8213/acdc91}{{\em Astrophys. J. Lett.}
  {\bfseries 951} no.~1, (2023) L11},
  \href{http://arxiv.org/abs/2306.16219}{{\ttfamily arXiv:2306.16219
  [astro-ph.HE]}}.

\bibitem{Assadullahi:2009nf}
H.~Assadullahi and D.~Wands, ``{Gravitational waves from an early matter
  era},'' \href{http://dx.doi.org/10.1103/PhysRevD.79.083511}{{\em Phys. Rev.
  D} {\bfseries 79} (2009) 083511},
  \href{http://arxiv.org/abs/0901.0989}{{\ttfamily arXiv:0901.0989
  [astro-ph.CO]}}.

\bibitem{Alabidi:2013lya}
L.~Alabidi, K.~Kohri, M.~Sasaki, and Y.~Sendouda, ``{Observable induced
  gravitational waves from an early matter phase},''
  \href{http://dx.doi.org/10.1088/1475-7516/2013/05/033}{{\em JCAP} {\bfseries
  05} (2013) 033}, \href{http://arxiv.org/abs/1303.4519}{{\ttfamily
  arXiv:1303.4519 [astro-ph.CO]}}.

\bibitem{Inomata:2019zqy}
K.~Inomata, K.~Kohri, T.~Nakama, and T.~Terada, ``{Gravitational Waves Induced
  by Scalar Perturbations during a Gradual Transition from an Early Matter Era
  to the Radiation Era},''
  \href{http://dx.doi.org/10.1088/1475-7516/2019/10/071}{{\em JCAP} {\bfseries
  10} (2019) 071}, \href{http://arxiv.org/abs/1904.12878}{{\ttfamily
  arXiv:1904.12878 [astro-ph.CO]}}. [Erratum: JCAP 08, E01 (2023)].

\bibitem{Inomata:2019ivs}
K.~Inomata, K.~Kohri, T.~Nakama, and T.~Terada, ``{Enhancement of Gravitational
  Waves Induced by Scalar Perturbations due to a Sudden Transition from an
  Early Matter Era to the Radiation Era},''
  \href{http://dx.doi.org/10.1103/PhysRevD.108.049901}{{\em Phys. Rev. D}
  {\bfseries 100} (2019) 043532},
  \href{http://arxiv.org/abs/1904.12879}{{\ttfamily arXiv:1904.12879
  [astro-ph.CO]}}. [Erratum: Phys.Rev.D 108, 049901 (2023)].

\bibitem{Pearce:2023kxp}
M.~Pearce, L.~Pearce, G.~White, and C.~Balazs, ``{Gravitational wave signals
  from early matter domination: interpolating between fast and slow
  transitions},'' \href{http://dx.doi.org/10.1088/1475-7516/2024/06/021}{{\em
  JCAP} {\bfseries 06} (2024) 021},
  \href{http://arxiv.org/abs/2311.12340}{{\ttfamily arXiv:2311.12340
  [astro-ph.CO]}}.

\bibitem{Domenech:2020ssp}
G.~Dom\`enech, C.~Lin, and M.~Sasaki, ``{Gravitational wave constraints on the
  primordial black hole dominated early universe},''
  \href{http://dx.doi.org/10.1088/1475-7516/2021/11/E01}{{\em JCAP} {\bfseries
  04} (2021) 062}, \href{http://arxiv.org/abs/2012.08151}{{\ttfamily
  arXiv:2012.08151 [gr-qc]}}. [Erratum: JCAP 11, E01 (2021)].

\bibitem{Gurian:2021rfv}
J.~Gurian, D.~Jeong, J.-c. Hwang, and H.~Noh, ``{Gauge-invariant tensor
  perturbations induced from baryon-CDM relative velocity and the B-mode
  polarization of the CMB},''
  \href{http://dx.doi.org/10.1103/PhysRevD.104.083534}{{\em Phys. Rev. D}
  {\bfseries 104} no.~8, (2021) 083534},
  \href{http://arxiv.org/abs/2104.03330}{{\ttfamily arXiv:2104.03330
  [astro-ph.CO]}}.

\bibitem{Malik:2008im}
K.~A. Malik and D.~Wands, ``{Cosmological perturbations},''
  \href{http://dx.doi.org/10.1016/j.physrep.2009.03.001}{{\em Phys. Rept.}
  {\bfseries 475} (2009) 1--51},
  \href{http://arxiv.org/abs/0809.4944}{{\ttfamily arXiv:0809.4944
  [astro-ph]}}.

\bibitem{Starobinsky:1986fx}
A.~A. Starobinsky, ``{STOCHASTIC DE SITTER (INFLATIONARY) STAGE IN THE EARLY
  UNIVERSE},'' \href{http://dx.doi.org/10.1007/3-540-16452-9_6}{{\em Lect.
  Notes Phys.} {\bfseries 246} (1986) 107--126}.

\bibitem{Starobinsky:1994bd}
A.~A. Starobinsky and J.~Yokoyama, ``{Equilibrium state of a selfinteracting
  scalar field in the De Sitter background},''
  \href{http://dx.doi.org/10.1103/PhysRevD.50.6357}{{\em Phys. Rev. D}
  {\bfseries 50} (1994) 6357--6368},
  \href{http://arxiv.org/abs/astro-ph/9407016}{{\ttfamily
  arXiv:astro-ph/9407016}}.

\bibitem{Linde:1996gt}
A.~D. Linde and V.~F. Mukhanov, ``{Nongaussian isocurvature perturbations from
  inflation},'' \href{http://dx.doi.org/10.1103/PhysRevD.56.R535}{{\em Phys.
  Rev. D} {\bfseries 56} (1997) R535--R539},
  \href{http://arxiv.org/abs/astro-ph/9610219}{{\ttfamily
  arXiv:astro-ph/9610219}}.

\bibitem{Sasaki:1987gy}
M.~Sasaki, Y.~Nambu, and K.-i. Nakao, ``{Classical Behavior of a Scalar Field
  in the Inflationary Universe},''
  \href{http://dx.doi.org/10.1016/0550-3213(88)90132-0}{{\em Nucl. Phys. B}
  {\bfseries 308} (1988) 868--884}.

\bibitem{Nambu:1987ef}
Y.~Nambu and M.~Sasaki, ``{Stochastic Stage of an Inflationary Universe
  Model},'' \href{http://dx.doi.org/10.1016/0370-2693(88)90974-4}{{\em Phys.
  Lett. B} {\bfseries 205} (1988) 441--446}.

\bibitem{Graham:2018jyp}
P.~W. Graham and A.~Scherlis, ``{Stochastic axion scenario},''
  \href{http://dx.doi.org/10.1103/PhysRevD.98.035017}{{\em Phys. Rev. D}
  {\bfseries 98} no.~3, (2018) 035017},
  \href{http://arxiv.org/abs/1805.07362}{{\ttfamily arXiv:1805.07362
  [hep-ph]}}.

\bibitem{Markkanen:2019kpv}
T.~Markkanen, A.~Rajantie, S.~Stopyra, and T.~Tenkanen, ``{Scalar correlation
  functions in de Sitter space from the stochastic spectral expansion},''
  \href{http://dx.doi.org/10.1088/1475-7516/2019/08/001}{{\em JCAP} {\bfseries
  08} (2019) 001}, \href{http://arxiv.org/abs/1904.11917}{{\ttfamily
  arXiv:1904.11917 [gr-qc]}}.

\bibitem{Hertzberg:2008wr}
M.~P. Hertzberg, M.~Tegmark, and F.~Wilczek, ``{Axion Cosmology and the Energy
  Scale of Inflation},''
  \href{http://dx.doi.org/10.1103/PhysRevD.78.083507}{{\em Phys. Rev. D}
  {\bfseries 78} (2008) 083507},
  \href{http://arxiv.org/abs/0807.1726}{{\ttfamily arXiv:0807.1726
  [astro-ph]}}.

\bibitem{Lyth:2001nq}
D.~H. Lyth and D.~Wands, ``{Generating the curvature perturbation without an
  inflaton},'' \href{http://dx.doi.org/10.1016/S0370-2693(01)01366-1}{{\em
  Phys. Lett. B} {\bfseries 524} (2002) 5--14},
  \href{http://arxiv.org/abs/hep-ph/0110002}{{\ttfamily arXiv:hep-ph/0110002}}.

\bibitem{Ando:2017veq}
K.~Ando, K.~Inomata, M.~Kawasaki, K.~Mukaida, and T.~T. Yanagida, ``{Primordial
  black holes for the LIGO events in the axionlike curvaton model},''
  \href{http://dx.doi.org/10.1103/PhysRevD.97.123512}{{\em Phys. Rev. D}
  {\bfseries 97} no.~12, (2018) 123512},
  \href{http://arxiv.org/abs/1711.08956}{{\ttfamily arXiv:1711.08956
  [astro-ph.CO]}}.

\bibitem{Kawasaki:2021ycf}
M.~Kawasaki and H.~Nakatsuka, ``{Gravitational waves from type II axion-like
  curvaton model and its implication for NANOGrav result},''
  \href{http://dx.doi.org/10.1088/1475-7516/2021/05/023}{{\em JCAP} {\bfseries
  05} (2021) 023}, \href{http://arxiv.org/abs/2101.11244}{{\ttfamily
  arXiv:2101.11244 [astro-ph.CO]}}.

\bibitem{Allali:2022yvx}
I.~J. Allali, M.~P. Hertzberg, and Y.~Lyu, ``{Altered axion abundance from a
  dynamical Peccei-Quinn scale},''
  \href{http://dx.doi.org/10.1103/PhysRevD.105.123517}{{\em Phys. Rev. D}
  {\bfseries 105} no.~12, (2022) 123517},
  \href{http://arxiv.org/abs/2203.15817}{{\ttfamily arXiv:2203.15817
  [hep-ph]}}.

\bibitem{Liddle:2003as}
A.~R. Liddle and S.~M. Leach, ``{How long before the end of inflation were
  observable perturbations produced?},''
  \href{http://dx.doi.org/10.1103/PhysRevD.68.103503}{{\em Phys. Rev. D}
  {\bfseries 68} (2003) 103503},
  \href{http://arxiv.org/abs/astro-ph/0305263}{{\ttfamily
  arXiv:astro-ph/0305263}}.

\bibitem{Dodelson:2003vq}
S.~Dodelson and L.~Hui, ``{A Horizon ratio bound for inflationary
  fluctuations},'' \href{http://dx.doi.org/10.1103/PhysRevLett.91.131301}{{\em
  Phys. Rev. Lett.} {\bfseries 91} (2003) 131301},
  \href{http://arxiv.org/abs/astro-ph/0305113}{{\ttfamily
  arXiv:astro-ph/0305113}}.

\bibitem{Podolsky:2005bw}
D.~I. Podolsky, G.~N. Felder, L.~Kofman, and M.~Peloso, ``{Equation of state
  and beginning of thermalization after preheating},''
  \href{http://dx.doi.org/10.1103/PhysRevD.73.023501}{{\em Phys. Rev. D}
  {\bfseries 73} (2006) 023501},
  \href{http://arxiv.org/abs/hep-ph/0507096}{{\ttfamily arXiv:hep-ph/0507096}}.

\bibitem{Munoz:2014eqa}
J.~B. Munoz and M.~Kamionkowski, ``{Equation-of-State Parameter for
  Reheating},'' \href{http://dx.doi.org/10.1103/PhysRevD.91.043521}{{\em Phys.
  Rev. D} {\bfseries 91} no.~4, (2015) 043521},
  \href{http://arxiv.org/abs/1412.0656}{{\ttfamily arXiv:1412.0656
  [astro-ph.CO]}}.

\bibitem{Lozanov:2016hid}
K.~D. Lozanov and M.~A. Amin, ``{Equation of State and Duration to Radiation
  Domination after Inflation},''
  \href{http://dx.doi.org/10.1103/PhysRevLett.119.061301}{{\em Phys. Rev.
  Lett.} {\bfseries 119} no.~6, (2017) 061301},
  \href{http://arxiv.org/abs/1608.01213}{{\ttfamily arXiv:1608.01213
  [astro-ph.CO]}}.

\bibitem{Maity:2018qhi}
D.~Maity and P.~Saha, ``{(P)reheating after minimal Plateau Inflation and
  constraints from CMB},''
  \href{http://dx.doi.org/10.1088/1475-7516/2019/07/018}{{\em JCAP} {\bfseries
  07} (2019) 018}, \href{http://arxiv.org/abs/1811.11173}{{\ttfamily
  arXiv:1811.11173 [astro-ph.CO]}}.

\bibitem{Antusch:2020iyq}
S.~Antusch, D.~G. Figueroa, K.~Marschall, and F.~Torrenti, ``{Energy
  distribution and equation of state of the early Universe: matching the end of
  inflation and the onset of radiation domination},''
  \href{http://dx.doi.org/10.1016/j.physletb.2020.135888}{{\em Phys. Lett. B}
  {\bfseries 811} (2020) 135888},
  \href{http://arxiv.org/abs/2005.07563}{{\ttfamily arXiv:2005.07563
  [astro-ph.CO]}}.

\bibitem{Allahverdi:2010xz}
R.~Allahverdi, R.~Brandenberger, F.-Y. Cyr-Racine, and A.~Mazumdar,
  ``{Reheating in Inflationary Cosmology: Theory and Applications},''
  \href{http://dx.doi.org/10.1146/annurev.nucl.012809.104511}{{\em Ann. Rev.
  Nucl. Part. Sci.} {\bfseries 60} (2010) 27--51},
  \href{http://arxiv.org/abs/1001.2600}{{\ttfamily arXiv:1001.2600 [hep-th]}}.

\bibitem{Abbott:1982hn}
L.~F. Abbott, E.~Farhi, and M.~B. Wise, ``{Particle Production in the New
  Inflationary Cosmology},''
  \href{http://dx.doi.org/10.1016/0370-2693(82)90867-X}{{\em Phys. Lett. B}
  {\bfseries 117} (1982) 29}.

\bibitem{Dolgov:1982th}
A.~D. Dolgov and A.~D. Linde, ``{Baryon Asymmetry in Inflationary Universe},''
  \href{http://dx.doi.org/10.1016/0370-2693(82)90292-1}{{\em Phys. Lett. B}
  {\bfseries 116} (1982) 329}.

\bibitem{Albrecht:1982mp}
A.~Albrecht, P.~J. Steinhardt, M.~S. Turner, and F.~Wilczek, ``{Reheating an
  Inflationary Universe},''
  \href{http://dx.doi.org/10.1103/PhysRevLett.48.1437}{{\em Phys. Rev. Lett.}
  {\bfseries 48} (1982) 1437}.

\bibitem{BICEP:2021xfz}
{\bfseries BICEP, Keck} Collaboration, P.~A.~R. Ade {\em et~al.}, ``{Improved
  Constraints on Primordial Gravitational Waves using Planck, WMAP, and
  BICEP/Keck Observations through the 2018 Observing Season},''
  \href{http://dx.doi.org/10.1103/PhysRevLett.127.151301}{{\em Phys. Rev.
  Lett.} {\bfseries 127} no.~15, (2021) 151301},
  \href{http://arxiv.org/abs/2110.00483}{{\ttfamily arXiv:2110.00483
  [astro-ph.CO]}}.

\bibitem{Cyr:2023pgw}
B.~Cyr, T.~Kite, J.~Chluba, J.~C. Hill, D.~Jeong, S.~K. Acharya, B.~Bolliet,
  and S.~P. Patil, ``{Disentangling the primordial nature of stochastic
  gravitational wave backgrounds with CMB spectral distortions},''
  \href{http://dx.doi.org/10.1093/mnras/stad3861}{{\em Mon. Not. Roy. Astron.
  Soc.} {\bfseries 528} no.~1, (2024) 883--897},
  \href{http://arxiv.org/abs/2309.02366}{{\ttfamily arXiv:2309.02366
  [astro-ph.CO]}}.

\bibitem{Carr:2020gox}
B.~Carr, K.~Kohri, Y.~Sendouda, and J.~Yokoyama, ``{Constraints on primordial
  black holes},'' \href{http://dx.doi.org/10.1088/1361-6633/ac1e31}{{\em Rept.
  Prog. Phys.} {\bfseries 84} no.~11, (2021) 116902},
  \href{http://arxiv.org/abs/2002.12778}{{\ttfamily arXiv:2002.12778
  [astro-ph.CO]}}.

\bibitem{Campeti:2020xwn}
P.~Campeti, E.~Komatsu, D.~Poletti, and C.~Baccigalupi, ``{Measuring the
  spectrum of primordial gravitational waves with CMB, PTA and Laser
  Interferometers},''
  \href{http://dx.doi.org/10.1088/1475-7516/2021/01/012}{{\em JCAP} {\bfseries
  01} (2021) 012}, \href{http://arxiv.org/abs/2007.04241}{{\ttfamily
  arXiv:2007.04241 [astro-ph.CO]}}.

\bibitem{Fedderke:2021kuy}
M.~A. Fedderke, P.~W. Graham, and S.~Rajendran, ``{Asteroids for
  \ensuremath{\mu}Hz gravitational-wave detection},''
  \href{http://dx.doi.org/10.1103/PhysRevD.105.103018}{{\em Phys. Rev. D}
  {\bfseries 105} no.~10, (2022) 103018},
  \href{http://arxiv.org/abs/2112.11431}{{\ttfamily arXiv:2112.11431 [gr-qc]}}.

\bibitem{Braglia:2021fxn}
M.~Braglia and S.~Kuroyanagi, ``{Probing prerecombination physics by the
  cross-correlation of stochastic gravitational waves and CMB anisotropies},''
  \href{http://dx.doi.org/10.1103/PhysRevD.104.123547}{{\em Phys. Rev. D}
  {\bfseries 104} no.~12, (2021) 123547},
  \href{http://arxiv.org/abs/2106.03786}{{\ttfamily arXiv:2106.03786
  [astro-ph.CO]}}.

\bibitem{Lu:2024yuo}
Z.~Lu, L.-T. Wang, and H.~Xiao, ``{A New Probe of $\mu$Hz Gravitational Waves
  with FRB Timing},'' \href{http://arxiv.org/abs/2407.12920}{{\ttfamily
  arXiv:2407.12920 [gr-qc]}}.

\bibitem{Sesana:2019vho}
A.~Sesana {\em et~al.}, ``{Unveiling the gravitational universe at $\mu$-Hz
  frequencies},'' \href{http://dx.doi.org/10.1007/s10686-021-09709-9}{{\em
  Exper. Astron.} {\bfseries 51} no.~3, (2021) 1333--1383},
  \href{http://arxiv.org/abs/1908.11391}{{\ttfamily arXiv:1908.11391
  [astro-ph.IM]}}.

\bibitem{Adshead:2021hnm}
P.~Adshead, K.~D. Lozanov, and Z.~J. Weiner, ``{Non-Gaussianity and the induced
  gravitational wave background},''
  \href{http://dx.doi.org/10.1088/1475-7516/2021/10/080}{{\em JCAP} {\bfseries
  10} (2021) 080}, \href{http://arxiv.org/abs/2105.01659}{{\ttfamily
  arXiv:2105.01659 [astro-ph.CO]}}.

\bibitem{Garcia-Saenz:2022tzu}
S.~Garcia-Saenz, L.~Pinol, S.~Renaux-Petel, and D.~Werth, ``{No-go theorem for
  scalar-trispectrum-induced gravitational waves},''
  \href{http://dx.doi.org/10.1088/1475-7516/2023/03/057}{{\em JCAP} {\bfseries
  03} (2023) 057}, \href{http://arxiv.org/abs/2207.14267}{{\ttfamily
  arXiv:2207.14267 [astro-ph.CO]}}.

\bibitem{Lyth:2002my}
D.~H. Lyth, C.~Ungarelli, and D.~Wands, ``{The Primordial density perturbation
  in the curvaton scenario},''
  \href{http://dx.doi.org/10.1103/PhysRevD.67.023503}{{\em Phys. Rev. D}
  {\bfseries 67} (2003) 023503},
  \href{http://arxiv.org/abs/astro-ph/0208055}{{\ttfamily
  arXiv:astro-ph/0208055}}.

\bibitem{Bartolo:2003jx}
N.~Bartolo, S.~Matarrese, and A.~Riotto, ``{On nonGaussianity in the curvaton
  scenario},'' \href{http://dx.doi.org/10.1103/PhysRevD.69.043503}{{\em Phys.
  Rev. D} {\bfseries 69} (2004) 043503},
  \href{http://arxiv.org/abs/hep-ph/0309033}{{\ttfamily arXiv:hep-ph/0309033}}.

\bibitem{Sasaki:2006kq}
M.~Sasaki, J.~Valiviita, and D.~Wands, ``{Non-Gaussianity of the primordial
  perturbation in the curvaton model},''
  \href{http://dx.doi.org/10.1103/PhysRevD.74.103003}{{\em Phys. Rev. D}
  {\bfseries 74} (2006) 103003},
  \href{http://arxiv.org/abs/astro-ph/0607627}{{\ttfamily
  arXiv:astro-ph/0607627}}.

\bibitem{Domenech:2021and}
G.~Dom\`enech, S.~Passaglia, and S.~Renaux-Petel, ``{Gravitational waves from
  dark matter isocurvature},''
  \href{http://dx.doi.org/10.1088/1475-7516/2022/03/023}{{\em JCAP} {\bfseries
  03} no.~03, (2022) 023}, \href{http://arxiv.org/abs/2112.10163}{{\ttfamily
  arXiv:2112.10163 [astro-ph.CO]}}.

\bibitem{Domenech:2023jve}
G.~Dom\`enech, ``{Cosmological gravitational waves from isocurvature
  fluctuations},'' \href{http://dx.doi.org/10.1007/s43673-023-00109-z}{{\em
  AAPPS Bull.} {\bfseries 34} no.~1, (2024) 4},
  \href{http://arxiv.org/abs/2311.02065}{{\ttfamily arXiv:2311.02065 [gr-qc]}}.

\bibitem{Dalianis:2020gup}
I.~Dalianis and C.~Kouvaris, ``{Gravitational waves from density perturbations
  in an early matter domination era},''
  \href{http://dx.doi.org/10.1088/1475-7516/2021/07/046}{{\em JCAP} {\bfseries
  07} (2021) 046}, \href{http://arxiv.org/abs/2012.09255}{{\ttfamily
  arXiv:2012.09255 [astro-ph.CO]}}.

\bibitem{Eggemeier:2022gyo}
B.~Eggemeier, J.~C. Niemeyer, K.~Jedamzik, and R.~Easther, ``{Stochastic
  gravitational waves from postinflationary structure formation},''
  \href{http://dx.doi.org/10.1103/PhysRevD.107.043503}{{\em Phys. Rev. D}
  {\bfseries 107} no.~4, (2023) 043503},
  \href{http://arxiv.org/abs/2212.00425}{{\ttfamily arXiv:2212.00425
  [astro-ph.CO]}}.

\bibitem{Fernandez:2023ddy}
N.~Fernandez, J.~W. Foster, B.~Lillard, and J.~Shelton, ``{Stochastic
  Gravitational Waves from Early Structure Formation},''
  \href{http://dx.doi.org/10.1103/PhysRevLett.133.111002}{{\em Phys. Rev.
  Lett.} {\bfseries 133} no.~11, (2024) 111002},
  \href{http://arxiv.org/abs/2312.12499}{{\ttfamily arXiv:2312.12499
  [astro-ph.CO]}}.

\bibitem{Dalianis:2024kjr}
I.~Dalianis and C.~Kouvaris, ``{Gravitational Waves from Collapse of
  Pressureless Matter in the Early Universe},''
  \href{http://arxiv.org/abs/2403.15126}{{\ttfamily arXiv:2403.15126
  [astro-ph.CO]}}.

\bibitem{2016ppap.book...87K}
T.~{Kluyver}, B.~{Ragan-Kelley}, F.~{P{\'e}rez}, B.~{Granger}, M.~{Bussonnier},
  J.~{Frederic}, K.~{Kelley}, J.~{Hamrick}, J.~{Grout}, S.~{Corlay},
  P.~{Ivanov}, D.~{Avila}, S.~{Abdalla}, C.~{Willing}, and {Jupyter Development
  Team}, \href{http://dx.doi.org/10.3233/978-1-61499-649-1-87}{``{Jupyter
  Notebooks{\textemdash}a publishing format for reproducible computational
  workflows},''} in {\em IOS Press}, pp.~87--90.
\newblock 2016.

\bibitem{harris2020array}
C.~R. Harris, K.~J. Millman, S.~J. van~der Walt, R.~Gommers, P.~Virtanen,
  D.~Cournapeau, E.~Wieser, J.~Taylor, S.~Berg, N.~J. Smith, R.~Kern, M.~Picus,
  S.~Hoyer, M.~H. van Kerkwijk, M.~Brett, A.~Haldane, J.~F. del R{\'{i}}o,
  M.~Wiebe, P.~Peterson, P.~G{\'{e}}rard-Marchant, K.~Sheppard, T.~Reddy,
  W.~Weckesser, H.~Abbasi, C.~Gohlke, and T.~E. Oliphant, ``Array programming
  with {NumPy},'' \href{http://dx.doi.org/10.1038/s41586-020-2649-2}{{\em
  Nature} {\bfseries 585} no.~7825, (Sept., 2020) 357--362}.
  \url{https://doi.org/10.1038/s41586-020-2649-2}.

\bibitem{2020SciPy-NMeth}
P.~Virtanen, R.~Gommers, T.~E. Oliphant, M.~Haberland, T.~Reddy, D.~Cournapeau,
  E.~Burovski, P.~Peterson, W.~Weckesser, J.~Bright, S.~J. {van der Walt},
  M.~Brett, J.~Wilson, K.~J. Millman, N.~Mayorov, A.~R.~J. Nelson, E.~Jones,
  R.~Kern, E.~Larson, C.~J. Carey, {\.I}.~Polat, Y.~Feng, E.~W. Moore,
  J.~{VanderPlas}, D.~Laxalde, J.~Perktold, R.~Cimrman, I.~Henriksen, E.~A.
  Quintero, C.~R. Harris, A.~M. Archibald, A.~H. Ribeiro, F.~Pedregosa, P.~{van
  Mulbregt}, and {SciPy 1.0 Contributors}, ``{{SciPy} 1.0: Fundamental
  Algorithms for Scientific Computing in Python},''
  \href{http://dx.doi.org/10.1038/s41592-019-0686-2}{{\em Nature Methods}
  {\bfseries 17} (2020) 261--272}.

\bibitem{Hunter:2007}
J.~D. Hunter, ``Matplotlib: A 2d graphics environment,''
  \href{http://dx.doi.org/10.1109/MCSE.2007.55}{{\em Computing in Science \&
  Engineering} {\bfseries 9} no.~3, (2007) 90--95}.

\bibitem{Lepage:1977sw}
G.~P. Lepage, ``{A New Algorithm for Adaptive Multidimensional Integration},''
  \href{http://dx.doi.org/10.1016/0021-9991(78)90004-9}{{\em J. Comput. Phys.}
  {\bfseries 27} (1978) 192}.

\bibitem{Lepage:2020tgj}
G.~P. Lepage, ``{Adaptive multidimensional integration: VEGAS enhanced},''
  \href{http://dx.doi.org/10.1016/j.jcp.2021.110386}{{\em J. Comput. Phys.}
  {\bfseries 439} (2021) 110386},
  \href{http://arxiv.org/abs/2009.05112}{{\ttfamily arXiv:2009.05112
  [physics.comp-ph]}}.

\bibitem{jax2018github}
J.~Bradbury, R.~Frostig, P.~Hawkins, M.~J. Johnson, C.~Leary, D.~Maclaurin,
  G.~Necula, A.~Paszke, J.~Vander{P}las, S.~Wanderman-{M}ilne, and Q.~Zhang,
  ``{JAX}: composable transformations of {P}ython+{N}um{P}y programs,'' 2018.
\newblock \url{http://github.com/google/jax}.

\bibitem{kidger2021on}
P.~Kidger, {\em {O}n {N}eural {D}ifferential {E}quations}.
\newblock PhD thesis, University of Oxford, 2021.

\bibitem{optimistix2024}
J.~Rader, T.~Lyons, and P.~Kidger, ``Optimistix: modular optimisation in jax
  and equinox,'' {\em arXiv:2402.09983} (2024) .

\bibitem{Mathematica}
W.~R. Inc., ``Mathematica, {V}ersion 14.1.''
\newblock \url{https://www.wolfram.com/mathematica}. Champaign, IL, 2024.

\bibitem{Acquaviva:2002ud}
V.~Acquaviva, N.~Bartolo, S.~Matarrese, and A.~Riotto, ``{Second order
  cosmological perturbations from inflation},''
  \href{http://dx.doi.org/10.1016/S0550-3213(03)00550-9}{{\em Nucl. Phys. B}
  {\bfseries 667} (2003) 119--148},
  \href{http://arxiv.org/abs/astro-ph/0209156}{{\ttfamily
  arXiv:astro-ph/0209156}}.

\end{thebibliography}\endgroup
\end{document}